\documentclass[11pt, a4paper]{article}
\pdfoutput=1
\usepackage{jcappub}
\usepackage{hyperref}
\usepackage{color}
\usepackage{graphicx}
\usepackage{physics}
\usepackage{bm}
\usepackage{amssymb}
\usepackage{empheq} 
\usepackage{xfrac} 
\usepackage[capitalise]{cleveref}
\usepackage{hyperref}
\hypersetup{
    colorlinks=true,
    linkcolor=blue,
    citecolor=red,
    filecolor=magenta,      
    urlcolor=magenta,
}
\usepackage[all]{hypcap}

\newcommand{\dT}{\delta T}

\newcommand{\cMHD}[1]{\ensuremath{\chi\mathrm{MHD}}}
\newcommand{\CPI}[1]{\ensuremath{\mathrm{CPI}}}
\newcommand{\CME}[1]{\ensuremath{\mathrm{CME}}}
\newcommand{\CVE}[1]{\ensuremath{\mathrm{CVE}}}
\newcommand{\CSE}[1]{\ensuremath{\mathrm{CSE}}}
\newcommand{\CESE}[1]{\ensuremath{\mathrm{CESE}}}
\newcommand{\AVE}[1]{\ensuremath{\mathrm{AVE}}}

\definecolor{brown(web)}{rgb}{0.65, 0.16, 0.16}

\newcommand{\ba}[1]{\begin{align} #1 \end{align}}
\newcommand{\bes}[1]{\begin{equation}\begin{split} #1 \end{split}\end{equation}}
\newcommand{\bsa}[2]{\begin{subequations}\label{#1}\begin{align} #2 \end{align}\end{subequations}}
\newcommand{\com}{\;,}
\newcommand{\per}{\;.}

\newcommand{\nn}{\nonumber \\}

\newcommand{\fref}[1]{figure~\ref{#1}}
\newcommand{\Fref}[1]{Figure~\ref{#1}}

\newcommand{\sref}[1]{section~\ref{#1}}

\newcommand{\pref}[1]{(\ref{#1})}
\newcommand{\eref}[1]{eq.~(\ref{#1})}
\newcommand{\Eref}[1]{Eq.~(\ref{#1})}
\newcommand{\erefs}[2]{eqs.~(\ref{#1})~and~(\ref{#2})}
\newcommand{\Erefs}[2]{Eqs.~(\ref{#1})~and~(\ref{#2})}
\newcommand{\rref}[1]{ref.~\cite{#1}}
\newcommand{\rrefs}[1]{refs.~\cite{#1}}

\newcommand{\aref}[1]{appendix~\ref{#1}}

\newcommand{\etal}[1]{et al.}
\newcommand{\eg}[1]{e.g.}
\newcommand{\ie}[1]{i.e.}
\newcommand{\cf}[1]{cf.}

\newcommand{\ii}{\mathrm{i}}
\newcommand{\ee}{\mathrm{e}}
\newcommand{\xvec}{{\bm x}}
\newcommand{\yvec}{{\bm y}}
\newcommand{\kvec}{{\bm k}}
\newcommand{\qvec}{{\bm q}}

\newcommand{\uvec}{{\bm u}}
\newcommand{\Xvec}{{\bm X}}
\newcommand{\Yvec}{{\bm Y}}
\newcommand{\Avec}{{\bm A}}
\newcommand{\Bvec}{{\bm B}}
\newcommand{\Evec}{{\bm E}}
\newcommand{\Svec}{{\bm S}}
\newcommand{\gvec}{{\bm g}}
\newcommand{\evec}{\mathbf{e}}
\newcommand{\jvec}{{\bm j}}
\newcommand{\Jvec}{{\bm J}}
\newcommand{\dvec}{{\bm \nabla}}
\newcommand{\Ssf}{\bm{\mathsf{S}}}
\newcommand{\Tsf}{\bm{\mathsf{T}}}
\newcommand{\SM}[1]{\text{\sc sm}}

\newcommand{\GeV}{\ \mathrm{GeV}}
\newcommand{\TeV}{\ \mathrm{TeV}}

\usepackage{ulem}

\title{\begin{center}
Plasma heating during the \\ chiral plasma instability
\end{center}}

\author[a]{Balin Armstrong,}
\emailAdd{ba63@rice.edu}
\author[a]{~Andrew~J.~Long,}
\emailAdd{andrewjlong@rice.edu}
\author[b,c]{~K\'aroly Seller,}
\emailAdd{karoly.seller@uni-hamburg.de}
\author[b]{~G\"unter Sigl}
\emailAdd{guenter.sigl@desy.de}

\affiliation[a]{Department of Physics and Astronomy, Rice University, Houston, TX, 77005, U.S.A.}
\affiliation[b]{Universit\"at Hamburg, II Institut f\"ur Theoretische Physik, 22761 Hamburg, Germany}
\affiliation[c]{Institute for Theoretical Physics, ELTE E\"otv\"os Lor\'and University, 1117 Budapest, Hungary}

\abstract{
The presence of a chiral asymmetry in a relativistic plasma opens a tachyonic instability toward the growth of a helical magnetic field.  We study the transfer of energy from the chiral asymmetry into the magnetic field during the development of this chiral plasma instability.  We find that there is more energy stored in the initial chiral asymmetry than goes into growing magnetic field and that the excess energy is transferred to the thermal bath.  Consequently, we find that the chiral plasma instability is accompanied by a heating of the plasma, and the temperature increase is parametrically $\dT \sim \mu_5^2 / T$ if the ratio of chemical potential to temperature is small, i.e. $\mu_5/T \ll 1$.  We briefly remark on possible observable implications for early universe cosmology. 
}

\begin{document}

\maketitle

\section{Introduction}
\label{sec:intro}

We are interested in plasmas with relativistic constituents and nonzero chiral charge.  
To have a concrete system in mind, the reader is invited to consider an electron-positron plasma at a temperature $T \gg m_e \approx 0.511 \; \mathrm{MeV}$ with an excess of right-chiral electrons/positrons over left-chiral electrons/positrons, but our work applies more generally to plasmas with other particle content.  
Relativistic chiral plasmas have been studied in the context of heavy ion collisions~\cite{Kharzeev:2013ffa,Kharzeev:2015znc,Koch:2016pzl,Zhao:2019hta,Kharzeev:2024zzm,Li:2025yxx}, compact stars~\cite{Boyarsky:2012ex,Dvornikov:2014cba,Grabowska:2014efa,Dvornikov:2014uza,Dvornikov:2015ena,Sigl:2015xva,Yamamoto:2015gzz,Kaplan:2016drz,Matsumoto:2022lyb}, and cosmology (see references below).  
In early-universe cosmology the primordial plasma does not need to possess a chiral charge in order to explain the observed Universe, but in many models of baryogenesis a chiral asymmetry arises alongside the baryon asymmetry, which is needed to explain the cosmological excess of matter over antimatter~\cite{Sakharov:1967dj}. 
Therefore it's reasonable to suppose that the primordial plasma may have sustained a nonzero chiral charge, and the implications of this hypothesis have been studied extensively. 

A relativistic chiral plasma is predicted to exhibit the chiral plasma instability (\CPI{}) whereby the chiral charge is depleted and fluctuations of the electromagnetic field are amplified, leading to the growth of a helical magnetic field with long-range spatial coherence~\cite{Joyce:1997uy,Frohlich:2000en}.  
Searches are underway for evidence of chiral effects in the quark-gluon plasma that arises at heavy ion collision experiments~\cite{Voloshin:2010ut,CMS:2016wfo,CMS:2017lrw,ALICE:2017sss,STAR:2021mii,Li:2025yxx}.  
In cosmology the \CPI{} offers a possible explanation for the origin of cosmological magnetism~\cite{Joyce:1997uy,Frohlich:2000en,Boyarsky:2011uy} (\ie{}, magnetogenesis), since the primordial magnetic field induced by the instability may survive in the universe today as a cosmological relic~\cite{Durrer:2013pga}.  
Our understanding of the \CPI{} in cosmology is informed by a combination of analytical calculations and numerical simulations~\cite{Brandenburg:1996fc,Boyarsky:2011uy,Dvornikov:2011ey,Boyarsky:2012ex,Tashiro:2012mf,Dvornikov:2013bca,Semikoz:2013xkc,Sigl:2015xva,Buividovich:2015jfa,Pavlovic:2016gac,Rogachevskii:2017uyc,Brandenburg:2017rcb,Anand:2018mgf,Brandenburg:2021aln,Kamada:2022nyt,Brandenburg:2023aco,Gurgenidze:2025lpt}.
The \CPI{} is a consequence of the chiral magnetic effect (\CME{})~\cite{Vilenkin:1980fu}, which is itself a manifestation of quantum field theoretic anomalous current nonconservation~\cite{Adler:1969gk,Bell:1969ts}.  
There is evidence for the \CME{} in measurements of condensed matter systems~\cite{Li:2014bha}, but it has not yet been observed in a high-energy environment. 
The \CPI{} therefore offers a rare opportunity to probe the physics of quantum anomalies in fundamental physics. 

This work is motivated by our desire to understand how energy is transferred between the fluid and the electromagnetic field as the \CPI{} develops.  
We build upon earlier studies that also addressed the question of energy balance~\cite{Dvornikov:2015ena,Sigl:2015xva,Long:2016uez,Kaplan:2016drz,Pavlovic:2018jiz}.\footnote{The authors of \rrefs{Sigl:2015xva,Kaplan:2016drz} raise criticisms of the arguments presented in \rref{Dvornikov:2015ena}, which is anyway irrelevant to our work as it focuses on neutron star environment rather than early universe.}  
The issue at hand can be understood as follows.  
If we denote the plasma temperature by $T$ and if we quantify the chiral charge with chemical potential $\mu_5$ (assumed to be $\ll T$), then the fluid's internal energy has two contributions: $\rho_T \sim T^4$ and $\rho_5 \sim \mu_5^2 T^2$.  
As the \CPI{} develops, the chiral charge is depleted and $\rho_5$ decreases toward zero.  
At the same time, a helical magnetic field develops, and the electromagnetic energy grows with $\rho_\mathrm{em} \sim B^2$.  
Are these changes precisely compensated, \ie{} $\Delta \rho_5 + \Delta \rho_\mathrm{em} = 0$?  Or if not, then where does the excess energy go?  


We seek to address these questions using the theory of chiral magnetohydrodynamics (\cMHD{})~\cite{Brandenburg:1996fc}.  
This classical field theory describes the evolution of the electromagnetic field, the fluid density, the fluid velocity, and the chiral charge in a relativistic chiral plasma.  
However, we find that the equations of motion of \cMHD{}, when expressed in their ``standard'' form that appears most frequently in the literature, are inadequate.  
We demonstrate how these equations are incompatible with exact energy conservation, and we explain why this behavior results from neglecting certain terms in the derivation, which are typically small but not exactly zero.  
We reintroduce the missing terms, and present an extended system of \cMHD{} equations of motion, which are compatible with local energy and momentum conservation.  
Using the corrected equations we investigate energy transfer between the fluid and the electromagnetic field.  
We find that an order one fraction of the chiral energy $\rho_5 \sim \mu_5^2 T^2$ is expended to create the magnetic field, and the remaining energy is released as heat, which causes the plasma temperature to rise slightly by $\dT \sim \mu_5^2 / T$.  


The article is organized as follows.  
In \sref{sec:CPI} we provide a short introduction to the \CPI{}, which highlights the basic mathematics of this tachyonic instability.  
In \sref{sec:chiral_MHD} we begin with a short review of \cMHD{}, and then we demonstrate how the equations of motion can be modified to accommodate exact energy and momentum conservation, which culminates in an extended system of equations.  
In \sref{sec:examples} we illustrate how the \CPI{} leads to heating by solving our extended \cMHD{} equations of motion, both analytically and numerically.  
Finally we conclude in \sref{sec:conc} with a short summary and some thoughts on how the heating might impact early universe cosmology.  
The article includes two appendices: \aref{app:RelativisticMHD} contains the equations of \cMHD{} in the nonrelativistic limit, and  \aref{app:chemical_potential_scaling} contains our analytical derivation of the late-time evolution.  

\section{Chiral plasma instability}
\label{sec:CPI}

In this section we briefly review how a helical magnetic field develops through the \CPI{} in a chiral plasma~\cite{Joyce:1997uy,Frohlich:2000en,Boyarsky:2011uy,Tashiro:2012mf}.  
For concreteness, consider a relativistic electron-positron plasma.  
The electrons, positrons, and photons are in thermal equilibrium at temperature $T$, with electric conductivity $\sigma$, magnetic diffusivity $\eta = 1/\sigma$, negligible fluid velocity $\uvec(\xvec,t) = 0$, and chiral chemical potential $\mu_5$.  
For $\mu_5 > 0$ the system contains an excess of right-chiral electrons and positrons over left-chiral electrons and positrons, and for $\mu_5 < 0$ the left-chiral particles are in excess.  
The magnetic induction equation is modified by the presence of the chiral asymmetry, and it takes the form 
\bes{
    \tfrac{\partial}{\partial t} \Bvec = \eta \nabla^2 \Bvec + \tfrac{2\alpha}{\pi} \mu_5 \eta \dvec \times \Bvec 
    \com
}
where $\alpha = e^2 / 4\pi$ is the electromagnetic fine structure constant.  
The magnetic field $\Bvec(\xvec,t)$ may be decomposed onto Fourier modes (wavevector $
\kvec$), and then further decomposed onto right and left-circular polarization modes (helicity $\pm$).  
Assuming that the plasma properties are homogeneous, the magnetic induction equation leads to 
\bes{
    \tfrac{\dd}{\dd t} B_{\kvec,\pm} = \Bigl( - \eta |\kvec|^2 \pm \tfrac{2\alpha}{\pi} \mu_5 \eta |\kvec| \Bigr) B_{\kvec,\pm} 
    \per
}
If the quantity in parenthesis is negative, then solutions decay exponentially toward zero.  
This is a consequence of the dissipative nature of the magnetic diffusivity $\eta$.  
Conversely, if the quantity in parenthesis is positive, then there is an instability that causes the field to grow.  
(The growth is exponential under the approximation that $\mu_5$ and $\eta$ remain constant, and in the next sections we'll discuss how the growth saturates as $\mu_5$ decreases.)  
This instability, known as the chiral plasma instability, is a tachyonic instability that impacts long-wavelength modes of the electromagnetic field.  
If $\mu_5 > 0$, then the right-circular polarization modes with $|\kvec| < k_\CPI{} = 2 \alpha |\mu_5| / \pi$ are unstable, and if $\mu_5 < 0$, then the left-circular polarization modes are unstable instead. 
The growing mode's solution is 
\bes{\label{eq:gen_ana_B_Field}
    B_{\kvec,+}(t) & = \mathrm{exp}\biggl[ \biggl( 1 - \biggl( \frac{|\kvec| - \tfrac{1}{2} k_\CPI{}}{\tfrac{1}{2} k_\CPI{}} \biggr)^2 \biggr) \biggl( \frac{t - t_i}{t_\CPI{}} \biggr) \biggr] \, B_{\kvec,+,i} 
    \com 
}
for $\mu_5 > 0$.  
The modes with $|\kvec| = \tfrac{1}{2} k_\CPI{}$ grow fastest on a time scale $t_\CPI{} = 4 / \eta k_\CPI{}^2 = \pi^2 / \eta \alpha^2 |\mu_5|^2$.  
As a consequence of the \CPI{}, thermal fluctuations of the magnetic field are amplified, leading to the development of a long-range helical magnetic field, which is dominated by right-circular polarization modes if the initial chiral asymmetry is positive ($\mu_5 > 0$), or left-circular polarization modes if it is negative.  

As the \CPI{} develops, the energy carried by the electromagnetic field increases.  
Our work is partly motivated by the question ``Where does this energy come from and where does it go?''.
In particular, is there an excess of chiral energy that is not used for magnetic field enhancement?

\section{Chiral magnetohydrodynamics}
\label{sec:chiral_MHD}

In this section, we present the equations of chiral magnetohydrodynamics (\cMHD{}) and examine their conservation properties. 
In their standard nonrelativistic form \cite{Brandenburg:1996fc,Brandenburg:2017rcb,Rogachevskii:2017uyc,Gurgenidze:2025lpt}, these equations do not satisfy exact conservation of energy, momentum, and chirality. 
This issue originates in the usual derivation, which neglects temperature variations and terms suppressed by the typically small chiral chemical potential. 
We identify the missing contributions that must be included in order to restore the exact conservation laws.

As pointed out in ref.~\cite{RoperPol:2025lgc}, the nonrelativistic limit taken in ref.~\cite{Brandenburg:1996fc} is not strictly consistent, as derivatives of the Lorentz $\gamma$ factor are discarded even though their effects are not generally subleading.  
We will not pursue this discrepancy here, since it is independent of the points we wish to make. 
Instead, we work with the standard nonrelativistic expressions of \rrefs{Brandenburg:1996fc,Brandenburg:2017rcb,Rogachevskii:2017uyc,Gurgenidze:2025lpt}, with the eventual aim of solving the equations in the limit of vanishing bulk velocity, where the differences between the various nonrelativistic limits disappear.

\subsection{Standard \cMHD{} equations}
\label{sec:eqns_of_chiMHD}

The electric and magnetic fields, $\Evec(\xvec,t)$ and $\Bvec(\xvec,t)$, satisfy Maxwell's equations:
\begin{subequations}\label{eq:Maxwell_eqn}
\begin{align}
    \label{eq:divE} 
    \dvec \cdot \Evec 
    & = Q
    \com 
    \\
    \label{eq:curlE} 
    \dvec \times \Evec
    & = - \tfrac{\partial}{\partial t} \Bvec 
    \com 
    \\
    \label{eq:divB} 
    \dvec \cdot \Bvec 
    & = 0 
    \com 
    \\
    \label{eq:curlB} 
    \dvec \times \Bvec 
    & = \tfrac{\partial}{\partial t} \Evec + \Jvec 
    \com 
\end{align}
\end{subequations}
where $Q(\xvec,t)$ is the electric charge density and $\Jvec(\xvec,t)$ is the electric current density.
In general, \erefs{eq:curlE}{eq:divB} are solved by $\Evec = - \dvec V - \tfrac{\partial}{\partial t} \Avec$ and $\Bvec = \dvec \times \Avec$, where $V(\xvec,t)$ and $\Avec(\xvec,t)$ are the electric scalar and magnetic vector potentials, respectively.
We shall work in the Weyl gauge, where the scalar potential is vanishing, $V(\xvec,t) = 0$. 

In the standard \cMHD{} framework, the following three general assumptions are made. 
(i) The plasma is locally electrically neutral\footnote{
One may question if this is appropriate to do so as one can find terms from the chiral magnetic effect and Ohm's law that generate an electric charge inhomogeneity. 
However, these terms are dependent on the fluid velocity and gradients of $\mu_5$ and $T$. 
As we are considering a non-relativistic fluid with initial fluctuations of these parameters to be small, we will assume that the generated electric charge inhomogeneity is negligible.}, \ie{}, we set $Q(\xvec,t)=0$.  
(ii) The electric current $\Jvec(\xvec,t)$ obeys the constitutive relation 
\ba{\label{eq:J_constitutive}
    \Jvec = \Jvec_\mathrm{Ohm} + \Jvec_\CME{} 
    \quad \text{with} \quad 
    \Jvec_\mathrm{Ohm} = \sigma (\Evec + \uvec \times \Bvec)
    \quad \text{and} \quad 
    \Jvec_\CME{} = \tfrac{2\alpha}{\pi} \mu_5 \Bvec
    \per
}
(iii)  The constituent particles of the plasma are relativistic, implying that the fluid pressure $p_\mathrm{fl}(\xvec,t)$ and the fluid energy density $\rho_\mathrm{fl}(\xvec,t)$ are related by the equation of state $p_\mathrm{fl} = \tfrac{1}{3} \rho_\mathrm{fl}$.
Furthermore, we employ a perturbative expansion in both the chiral chemical potential $\mu_5$ and the bulk velocity $\uvec$.
In the rest of this paper, if not indicated otherwise, we shall only keep the leading non-trivial terms in $\mu_5/T\ll 1$ and $|\uvec|\ll 1$.

The energy in a chiral plasma is shared between the electromagnetic field and the fluid.  
The total energy density $\rho_\mathrm{tot}(\xvec,t)$ is decomposed as 
\begin{subequations}
\label{eq:rho_def} 
\ba{\label{eq:rho_tot}
    \rho_\mathrm{tot} = \rho_\mathrm{em} + \rho_\mathrm{fl} + \rho_\mathrm{kin} 
    \com
}
where the three terms correspond to the energy density of the electromagnetic field, the internal energy density of the fluid, and the bulk kinetic energy density of the fluid, respectively.
The internal energy density is further decomposed as the sum of the thermal and chiral energy densities, 
\ba{
\label{eq:rho_fl}
    \rho_\mathrm{fl} = \rho_T + \rho_5 
    \per 
}
The various components are given by the following expressions 
\ba{
    \rho_\mathrm{em} & = \tfrac{1}{2} |\Evec|^2 + \tfrac{1}{2} |\Bvec|^2 \com \label{eq:rho_em} \\ 
    \rho_T & = \tfrac{\pi^2}{30} g_E T^4 \com \label{eq:rho_T} \\ 
    \rho_5 & = \tfrac{1}{2} \mu_5^2 T^2 + \tfrac{1}{4\pi^2} \mu_5^4 = \tfrac{1}{2} \mu_5^2 T^2 \bigl[ 1 + \mathcal{O}\bigl(\mu_5^2 / T^2 \bigr) \bigr] \com \label{eq:rho_5} \\ 
    \rho_\mathrm{kin} & = \tfrac{4}{3} \rho_\mathrm{fl} \tfrac{|\uvec|^2}{1-|\uvec|^2} = \tfrac{4}{3} \rho_\mathrm{fl} |\uvec|^2 \bigl[ 1 + \mathcal{O}(|\uvec|^2) \bigr] 
    \com
}
\end{subequations}
where $T(\xvec,t)$ is the plasma temperature and $g_E$ is the effective number of relativistic species constituting the plasma.
For $\rho_5$ and $\rho_\mathrm{kin}$, the exact formulae are given first and then their expansions to leading order in the small parameters. 

The observable parity-odd properties of the \cMHD{} system are characterized by the chiral number density $n_5(\xvec,t) \equiv n_R - n_L$ and the magnetic helicity density $h(\xvec,t)$. 
These densities are calculated as 
\begin{align}\label{eq:n5_and_h}
    n_5 
    = \tfrac{1}{3} \mu_5 T^2 + \tfrac{1}{3\pi^2} \mu_5^3  
    = \tfrac{1}{3}\mu_5 T^2 \bigl[ 1 + \mathcal{O}\bigl(\mu_5^2 / T^2 \bigr) \bigr] \qquad 
    \text{and} \qquad 
    h = \Avec \cdot \Bvec  
    \per 
\end{align}  
While the latter is not gauge invariant, its volume average $\tfrac{1}{V} \int_\mathcal{V} \dd^3 \xvec \, h(\xvec,t)$ is so, provided that $\Bvec(\xvec,t)$ is perpendicular to the surface bounding the volume $\mathcal{V}$ (usually taken to be the infinite volume limit). 
The chirality and helicity evolve according to~\cite{Joyce:1997uy} 
\bes{\label{eq:dn5dt_and_dhdt}
    \tfrac{\partial}{\partial t} n_5 + \dvec \cdot \jvec_5 & = \tfrac{2\alpha}{\pi} \Evec \cdot \Bvec - \Gamma_5 n_5 \\ 
    \tfrac{\partial}{\partial t} h & = - 2 \Evec \cdot \Bvec 
    - \dvec \cdot \bigl( \Evec \times \Avec \bigr)
    \per
}
The first equation is the Adler–Bell–Jackiw chiral anomaly with the chiral current density given by $\jvec_5(\xvec,t) = n_5 \uvec - D_5 \dvec n_5 $, where $D_5(\xvec,t)$ is the chiral diffusion coefficient.
An additional term was also introduced to account for scattering processes in the plasma that violate chirality at a rate $\Gamma_5(\xvec,t)$.
The second equation is an identity that follows from Maxwell's equations (in the Weyl gauge). 

The \cMHD{} equations of motion form a set of coupled partial differential equations for the dynamical fields of the system.
For nonrelativistic bulk velocities and a small chiral asymmetry, the approximation 
leads to the standard system of equations \cite{Brandenburg:1996fc,Brandenburg:2017rcb,Rogachevskii:2017uyc,RoperPol:2025lgc,Gurgenidze:2025lpt}:
\begin{subequations}\label{eq:cMHD}
\begin{align}
    \label{eq:dAdt} 
    \tfrac{\partial}{\partial t} \Avec 
    & = \uvec \times \Bvec 
    + \eta \bigl( \tfrac{2\alpha}{\pi} \mu_5 \, \Bvec - \Jvec \bigr) 
    \com 
    \\
    \label{eq:dmudt}
    \bigl( \tfrac{\partial}{\partial t} + \uvec \cdot \dvec \bigr) \bigl( \tfrac{2\alpha}{\pi} \mu_5 \bigr)   
    & = - \tfrac{2\alpha}{\pi} \mu_5 \dvec \cdot \uvec  
    + D_5 \nabla^2 \bigl( \tfrac{2\alpha}{\pi} \mu_5 \bigr) 
    \\ & \quad 
    - \lambda \eta \bigl( \tfrac{2\alpha}{\pi} \mu_5 \Bvec - \Jvec \bigr) \cdot \Bvec 
    - \Gamma_5 \bigl( \tfrac{2\alpha}{\pi} \mu_5 \bigr) + O(\mu_5^2) 
    \com\nonumber
    \\ 
    \label{eq:dudt}
    \rho_\mathrm{fl} \bigl( \tfrac{\partial}{\partial t} + \uvec \cdot \dvec \bigr) \uvec 
    & = 
    2 \dvec \cdot \bigl( \rho_\mathrm{fl} \nu \Ssf \bigr) 
    - \tfrac{1}{4} \dvec \rho_\mathrm{fl} 
    + \tfrac{1}{3} \uvec \dvec \cdot (\rho_\mathrm{fl} \uvec)
    \\ & \quad 
    + \tfrac{3}{4} \Jvec \times \Bvec 
    - \uvec \bigl[ \uvec \cdot (\Jvec \times \Bvec) + \eta |\Jvec|^2 \bigr] 
    \com\nonumber
    \\
    \label{eq:drhofldt}
    \tfrac{\partial}{\partial t} \rho_\mathrm{fl} 
    & = 
    - \tfrac{4}{3} \dvec \cdot (\rho_\mathrm{fl} \uvec)
    + \uvec \cdot (\Jvec\times\Bvec) 
    + \eta |\Jvec|^2
    \per 
\end{align}
\end{subequations}
The first equation is simply Maxwell's equation in the Weyl gauge with $\Jvec$ given by the constitutive relation in \eref{eq:J_constitutive}.
The second follows from the anomaly equation \eqref{eq:dn5dt_and_dhdt} after substitution of $n_5$ from \eref{eq:n5_and_h} in the $T={\rm constant}$ limit.
The remaining two are the spatial and temporal components of the energy-momentum continuity equation.
Furthermore, the traceless rate-of-strain tensor $\Ssf(\xvec,t)$ has components ${\sf S}_{ij} = ( \partial_j u_i + \partial_i u_j ) / 2 - \delta_{ij} \partial_k u_k/3$, while
$\nu$ is the kinematic viscosity and $\lambda = \tfrac{192}{T^2} (\tfrac{\alpha}{4\pi})^2$ is the chiral feedback parameter \cite{Brandenburg:2023aco,Gurgenidze:2025lpt}. 

\subsection{Failure of conservation laws}
\label{sec:energyconservation}

Using the \cMHD{} equations of motion, which appear in \erefs{eq:Maxwell_eqn}{eq:cMHD}, one may ask how the energy, momentum, and chirality of the system evolve with time.  
Here we demonstrate how the expected conservation laws are not obtained using the standard \cMHD{} equations. 

An expression for the total energy density $\rho_\mathrm{tot}$ appears in \eqref{eq:rho_def}.  
Its time derivative is 
\bes{
    \tfrac{\partial}{\partial t} \rho_\mathrm{tot} 
    & = \tfrac{\partial}{\partial t} \rho_\mathrm{fl} 
    + \tfrac{\partial}{\partial t} \rho_\mathrm{kin} 
    + \Evec \cdot \tfrac{\partial}{\partial t} \Evec 
    + \Bvec \cdot \tfrac{\partial}{\partial t} \Bvec 
    \per
}
The equations of motion in eqs.~\eqref{eq:curlE},~\eqref{eq:curlB},~and~\eqref{eq:drhofldt} are used to derive 
\bes{\label{eq:dt_rho_tot}
    \tfrac{\partial}{\partial t} \rho_\mathrm{tot} 
    + \dvec \cdot \Svec_\mathrm{tot}
    = \tfrac{2\alpha}{\pi} \eta \mu_5 \Bvec \cdot \Jvec 
    + \mathcal{O}(|\uvec|) 
    \neq 0
    \com
}
where $\Svec_\mathrm{tot} = \Svec_\mathrm{fl} + \Svec_\mathrm{em}$ is the total energy flux with $\Svec_\mathrm{fl} = \tfrac{4}{3} \rho_\mathrm{fl} \uvec$ and $\Svec_\mathrm{em} = \Evec \times \Bvec$ being the fluid and electromagnetic energy fluxes, respectively.
Since the right-hand side of \eref{eq:dt_rho_tot} is nonzero in general, this calculation demonstrates that the standard \cMHD{} equations are not compatible with local energy conservation.  
Note, however, that the measure of the violation is proportional to $\mu_5$ and $\eta$, both parameters that are assumed to be small.
The non-conservation of energy is due to the standard equations \eqref{eq:cMHD} not being fully self-consistent: some terms are dropped due the presumed smallness of the chiral chemical potential.

The total momentum density $\gvec_\mathrm{tot}(\xvec,t)$ is equal to the total energy flux, which implies the relation $\gvec_\mathrm{tot} = \tfrac{4}{3} \rho_\mathrm{fl} \uvec + \Evec \times \Bvec$.  
Its time derivative is 
\begin{equation}
    \label{eq:dgtotdt}
    \tfrac{\partial}{\partial t} \gvec_\mathrm{tot}
    = \tfrac{4}{3}\uvec \tfrac{\partial}{\partial t} \rho_\mathrm{fl} + \tfrac{4}{3} \rho_\mathrm{fl} \tfrac{\partial}{\partial t} \uvec + \tfrac{\partial}{\partial t} \Evec \times \Bvec + \Evec \times \tfrac{\partial}{\partial t} \Bvec 
    \per
\end{equation}
The equations of motion in eqs.~\eqref{eq:curlE},~\eqref{eq:curlB},~\eqref{eq:dudt},~and~\eqref{eq:drhofldt} are used to derive 
\begin{equation}
\label{eq:momentum_conservation_approx}
    \tfrac{\partial}{\partial t} \gvec_\mathrm{tot} + \dvec \cdot \Tsf_\mathrm{tot} = 0 
    \com 
\end{equation}
where $\Tsf_\mathrm{tot}(\xvec,t)=\Tsf_\mathrm{fl}(\xvec,t) + \Tsf_\mathrm{em}(\xvec,t)$ is the total energy-momentum 3-tensor with the fluid energy-momentum tensor $T^{ij}_\mathrm{fl} = \tfrac{1}{3}\rho_\mathrm{fl}(4u_i u_j + \delta_{ij})$ and the electromagnetic energy tensor (\ie{}, Maxwell stress tensor) $T^{ij}_\mathrm{em} = \delta_{ij} \rho_\mathrm{em} - E_i E_j - B_i B_j$. 
Here we have neglected the viscosity term, which is $\propto \nu$ in \eref{eq:dudt}, but its inclusion is trivial and does not modify our arguments.  
\Eref{eq:momentum_conservation_approx} demonstrates that the equations of \cMHD{} do exhibit momentum conservation (in the absence of viscosity).   
However, we will find that \eref{eq:drhofldt} must be modified in order to enforce energy conservation, and this will disrupt momentum conservation, forcing us to also modify \eref{eq:dudt}.
We will address the issue of energy non-conservation and derive the modified continuity equation in the next section.

In addition to the non-conservation of energy (and momentum), one can further show that the equations of \cMHD{} in \eref{eq:cMHD} are inconsistent with the first law of thermodynamics.
Thermodynamical arguments can be applied in local comoving frames where $\uvec$ vanishes.  
The first law of thermodynamics then takes the form 
\begin{equation}
    \label{eq:thermo_fundamental_relation}
    \tfrac{\partial}{\partial t}\rho_\mathrm{fl} = T \tfrac{\partial}{\partial t} s_\mathrm{fl} + \mu_5 \tfrac{\partial}{\partial t} n_5\com
\end{equation}
where $s_\mathrm{fl}$ is the entropy density of the fluid.
\Eref{eq:thermo_fundamental_relation} suggests that for $\uvec=0$ the anomaly equation in \eref{eq:dmudt} and the energy continuity equation in \eref{eq:drhofldt} can be used to derive the generally non-zero entropy injection into the fluid.
In the local comoving frame, using the definition of the chiral number density in \eref{eq:n5_and_h}, one finds
\begin{subequations}
\label{eq:int_dot_rho_fl_dot_n5}
\begin{align}
    \label{eq:anomaly_eq_intV}
    &\eqref{eq:dmudt} \quad \rightarrow \quad \left.\tfrac{\partial}{\partial t}n_5\right|_{T=T_0} = \tfrac{2\alpha}{\pi} \Evec\cdot\Bvec  - \Gamma_5 n_5|_{T=T_0} + D_5\dvec^2 n_5|_{T=T_0}\com
    \\
    \label{eq:continuity_eq_intV}
    &\eqref{eq:drhofldt} \quad \rightarrow \quad \tfrac{\partial}{\partial t}\rho_{\rm fl} = \sigma|\Evec|^2 + \tfrac{2\alpha}{\pi} \eta \mu_5 \Bvec \cdot \Jvec|_{\uvec=0} + \mu_5 \tfrac{2\alpha}{\pi}\Evec\cdot\Bvec \per
\end{align}
\end{subequations}
In the first line, we set the temperature and the diffusion coefficient to be constants, as no derivatives of $T$ or $D_5$ appeared in \eref{eq:dmudt}.
We can combine \erefs{eq:anomaly_eq_intV}{eq:continuity_eq_intV} and write them in the form resembling the fundamental equation of thermodynamics given in \eref{eq:thermo_fundamental_relation}:
\begin{equation}
    \label{eq:energy_non_conservation}
    \tfrac{\partial}{\partial t}\rho_{\rm fl} = \big[\sigma|\Evec|^2+ \tfrac{2\alpha}{\pi} \eta \mu_5 \Bvec \cdot \Jvec|_{\uvec=0}\big] + \mu_5\left[\left.\tfrac{\partial}{\partial t}n_5\right|_{T=T_0} + \Gamma_5 n_5|_{T=T_0}  - D_5\dvec^2 n_5|_{T=T_0}\right]\per
\end{equation}
Consistency between the sides requires that $T=T_0$ is kept constant throughout.
Using \eref{eq:thermo_fundamental_relation} then gives
\begin{subequations}
\label{eq:Tds}
\begin{align}
    \label{eq:Tds1}
    T_0 \left.\tfrac{\partial}{\partial t} s_{\rm fl}\right|_{T=T_0}= \sigma|\Evec|^2+ \tfrac{2\alpha}{\pi} \eta \mu_5 \Bvec \cdot \Jvec|_{\uvec=0} + \mu_5\left[\Gamma_5n_5|_{T=T_0} - D_5\dvec^2 n_5|_{T=T_0}\right]& \\
    \label{eq:Tds2}
    = \sigma |\Evec|^2 + \left[\tfrac{\partial}{\partial t}\rho_{\rm tot}+\dvec\cdot\Svec_{\rm tot}\right]_{\uvec=0,T=T_0} + \mu_5\left[\Gamma_5n_5|_{T=T_0} - D_5\dvec^2 n_5|_{T=T_0}\right]& \com
\end{align}
\end{subequations}
where in the second equality we used \eref{eq:dt_rho_tot}.
We shall see in the following subsection that this identification is justified, \ie{}, this term vanishes if the energy continuity equation is satisfied.
The important point here is that $\sigma |\Evec|^2$, contrary to the rest of the terms on the right-hand side, is independent of the chemical potential, and thus it is present even if we set $\mu_5=0$.
We then note that if the chemical potential is vanishing, $s_{\rm fl}|_{T=T_0}\propto T_0^3$ is a constant and the left-hand side of \eref{eq:Tds} is zero. This, however, leads to a contradiction, as in general $\sigma |\Evec|^2\neq 0$.
The only way out is to allow the temperature to change, leading to additional terms proportional to the derivatives of $T$ in the anomaly equation \pref{eq:dmudt}.  
The authors of \rref{Rogachevskii:2017uyc} also noted that temperature should be treated as a field that varies across space and time, with its evolution determined by entropy injection.
They have not presented the temperature-derivative-dependent equations, however, as they argued that temperature gradients relax quickly, making the homogeneous treatment self-consistent.

\subsection{Restoring conservation laws}
\label{sec:restoring}

In the previous subsection, we pointed out two discrepancies in the standard \cMHD{} equations: (i) energy is not exactly conserved and (ii) energy conservation is incompatible with the anomaly equation at fixed temperature. 
Here we shall address both of these points. 

Both energy and momentum conservation may be expressed in a Lorentz covariant form through the energy-momentum continuity equation.  
For a plasma consisting of a charged fluid and electromagnetic field, the energy-momentum continuity equation is expressed as 
\begin{equation}
    \label{eq:continuity_eq}
    \partial_\mu (T^{\mu\nu}_\mathrm{fl}+T^{\mu\nu}_\mathrm{em})=0\com
\end{equation}
where $T^{\mu\nu}_{\rm a}$ are the stress-energy tensors of the respective components. 
The stress-energy tensor for an ultra-relativistic perfect fluid is 
\begin{equation}
    \label{eq:T_munu_fl}
    T^{\mu\nu}_\mathrm{fl}=\rho_\mathrm{fl}\left(\tfrac{4}{3}u^\mu u^\nu - \tfrac{1}{3} \eta^{\mu\nu}\right)\com
\end{equation}
where $\rho_\mathrm{fl}$ is the fluid's (Lorentz-invariant) internal energy density, $u^\mu = \gamma (1,\uvec)$ is the fluid's 4-velocity, and $\eta^{\mu\nu}={\rm diag}(1,-1,-1,-1)$ is the Minkowski metric tensor.
Similarly, the stress-energy tensor for the electromagnetic field is:
\begin{equation}
    \label{eq:T_munu_em}
    T^{\mu\nu}_\mathrm{em}=\tfrac{1}{4}\eta^{\mu\nu}F^{\rho\sigma}F_{\rho\sigma} - F^{\mu\sigma}\eta^{\nu\rho}F_{\rho\sigma}\com
\end{equation}
where $F_{\mu\nu}=\partial_\mu A_\nu-\partial_\nu A_\mu$ is the electromagnetic field-strength tensor.

If the fluid's motion remains nonrelativistic, then $\gamma \approx 1$.  
Setting $u^\mu = (1, \uvec)$ in \eref{eq:continuity_eq} leads to the following energy and momentum continuity equations:\footnote{
As pointed out in \rref{RoperPol:2025lgc}, the derivation should be done such that the limit $\gamma\to 1$ is only taken at the very end of the calculation.
This allows partial derivatives to act on the Lorentz $\gamma$-factors in \eqref{eq:T_munu_fl} and introduces further leading order terms missed by the naive calculation.
The resulting non-relativistic limit is slightly different from the equations presented here, nevertheless, it is unrelated to energy non-conservation or violation of thermodynamical principles, and thus is outside the focus of this work.
We also mention that the differences among the  evolution equations are explicitly dependent on the bulk motion, thus setting $|\uvec|= 0$ results in identical equations. 
In  \aref{app:RelativisticMHD}, we show what the equations of motion should be if we properly took the limit $\gamma\to 1$.
}
\begin{subequations}
\label{eq:cMHD_fix_rho}
\begin{align}
    \tfrac{\partial}{\partial t} \rho_\mathrm{fl} 
    & = \left[\tfrac{\partial}{\partial t} \rho_\mathrm{fl}\right]_{\rm approx.} 
    - \tfrac{2\alpha }{\pi}\eta\mu_5\Bvec \cdot \Jvec 
    \com 
    \\
    \rho_\mathrm{fl} \bigl( \tfrac{\partial}{\partial t} + \uvec \cdot \dvec \bigr) \uvec 
    & = \left[\rho_\mathrm{fl} \bigl( \tfrac{\partial}{\partial t} + \uvec \cdot \dvec \bigr) \uvec\right]_{\rm approx.} + (\tfrac{2\alpha }{\pi}\eta\mu_5\Bvec\cdot \Jvec) \, \uvec 
    \com
\end{align}
\end{subequations}
where the terms with the ``approx.'' subscripts are understood as
the right-hand sides of \erefs{eq:dudt}{eq:drhofldt}. 
Taking $u^0 = 1$ rather than $\gamma$ amounts to neglecting $\rho_\mathrm{kin}$ relative to $\rho_\mathrm{fl}$. 
With these modified equations of motion at hand, one can revisit the calculations in \sref{sec:energyconservation} that demonstrated the failure of energy conservation. 
Thanks to the additional terms in \eref{eq:cMHD_fix_rho}, one finds that both energy and momentum are now conserved in the sense that the right-hand side of \eref{eq:dt_rho_tot} becomes zero and that the right-hand side of \eref{eq:momentum_conservation_approx} remains zero.

We now turn to the anomaly equation.
In the previous subsection, we showed that for the \cMHD{} equations \pref{eq:cMHD}, the fundamental equation of thermodynamics is not satisfied unless temperature is allowed to change, which contradicted \eref{eq:dmudt}.
In light of the proper energy continuity equations given in \eref{eq:cMHD_fix_rho}, we recalculate the fundamental relation and find that the requirement for temperature change is upheld.
We find
\begin{equation}
    T_0 \left.\tfrac{\partial}{\partial t} s_{\rm fl}\right|_{T=T_0}
    = \sigma |\Evec|^2 + \mu_5\left(\Gamma_5n_5|_{T=T_0} - D_5\dvec^2 n_5|_{T=T_0}\right)\,,
\end{equation}
which again leads to the same contradiction as explained below \eref{eq:Tds}.
Allowing temperature to change formally leads to the same equation. 
In addition, we can identify the terms on the right-hand side as Ohmic heating and entropy injections due to chirality-changing particle interactions and diffusion, respectively.\footnote{
We have implicitly assumed that thermalization can occur on timescales much shorter than the CPI timescale, allowing temperature to be consistently defined for the plasma.
This is justified by noting that $t_\CPI{}\propto\sigma/k_\CPI{}^2$, whereas one could naively estimate the thermalization time for modes around the wavenumber $k_\CPI{}$ as $t_{\rm therm.}\propto1/k_\CPI{}\ll t_\CPI{}$.}

\subsection{Extended \cMHD{} equations}
\label{sec:extended}

After implementing the corrections described in the previous section, we obtain a new set of \cMHD{} equations of motion that replace those in \eqref{eq:cMHD}.  
These new equations govern the evolution of non-relativistic \cMHD{}, and they are consistent with energy conservation and the anomaly equation.  
Our equations can be expressed in several forms depending on what quantities are chosen as the dependent variables.  
In the most compact form, we have
\begin{subequations}
\begin{align}
    \label{eq:MHD_dA_dt}
    \tfrac{\partial}{\partial t}\Avec &= \uvec\times\Bvec + \eta\left(\tfrac{2\alpha}{\pi}\mu_5\Bvec -\Jvec\right)\com 
    \\
    \label{eq:MHD_dn5_dt}
    \left[\tfrac{\partial n_5}{\partial T}\right]\tfrac{\partial}{\partial t}T + \left[\tfrac{\partial n_5}{\partial \mu_5}\right]\tfrac{\partial}{\partial t}\mu_5 &= \tfrac{2\alpha}{\pi}\Evec\cdot\Bvec - \dvec\cdot\jvec_5 - \Gamma_5n_5\com
    \\
    \label{eq:MHD_drhofl_dt}
    \left[\tfrac{\partial \rho_\mathrm{fl}}{\partial T}\right]\tfrac{\partial}{\partial t}T + \left[\tfrac{\partial \rho_\mathrm{fl}}{\partial \mu_5}\right]\tfrac{\partial}{\partial t}\mu_5 &= \Evec\cdot\Jvec - \tfrac{4}{3}\dvec\cdot(\rho_\mathrm{fl}\uvec)
    \\
    \label{eq:MHD_Du_dt}
    \rho_\mathrm{fl}\left(\tfrac{\partial}{\partial t}+\uvec\cdot\dvec\right)\uvec &= 2\dvec\cdot(\rho_\mathrm{fl}\nu\Ssf)-\tfrac{1}{4} \dvec \rho_\mathrm{fl} + \tfrac{1}{3} \rho_\mathrm{fl} \uvec(\dvec\cdot\uvec) 
    + \tfrac{3}{4} \Jvec \times \Bvec -\uvec ( \Evec \cdot \Jvec )
    \per
\end{align}
\end{subequations}
The derivatives in square brackets may be evaluated using the expressions in \erefs{eq:rho_def}{eq:n5_and_h}.
For easier comparison with the standard \cMHD{} equations \eqref{eq:cMHD}, we substitute $n_5 \simeq \mu_5 T^2 / 3$ while retaining $\rho_\mathrm{fl}$.  
Highlighting the new terms in red, we have
\begin{subequations}\label{eq:cMHD_corrected}
\begin{align}
    \label{eq:cMHD_final_Maxwell}
    \tfrac{\partial}{\partial t} \Avec 
    & = \uvec \times \Bvec 
    + \eta \bigl( \tfrac{2\alpha}{\pi} \mu_5 \, \Bvec - \Jvec \bigr) 
    \com 
    \\
    \label{eq:cMHD_final_anomaly}
    \bigl( \tfrac{\partial}{\partial t} + \uvec \cdot \dvec \bigr) \bigl( \tfrac{2\alpha}{\pi} \mu_5 \bigr)   
    & = - \bigl( \tfrac{2\alpha}{\pi} \mu_5 \bigr) \dvec \cdot \uvec  
    + D_5 \nabla^2 \bigl( \tfrac{2\alpha}{\pi} \mu_5 \bigr) 
    \\ & \quad 
    - \lambda \eta \bigl( \tfrac{2\alpha}{\pi} \mu_5 \Bvec - \Jvec \bigr) \cdot \Bvec 
    - \Gamma_5 \bigl( \tfrac{2\alpha}{\pi} \mu_5 \bigr)\nonumber 
    \\ & \quad 
    - {\color{red}
    \tfrac{2}{T} \bigl( \tfrac{2\alpha}{\pi} \mu_5 \bigr) \bigl( \tfrac{\partial}{\partial t} + \uvec \cdot \dvec \bigr) T 
    }
    + {\color{red} 
    D_5 \tfrac{1}{T^2} \bigl( \tfrac{2\alpha}{\pi} \mu_5 \bigr) \nabla^2 \bigl( T^2\bigr)
    }
    \nn & \quad 
    + {\color{red} 
    D_5 \tfrac{4}{T} \bigl( \dvec T \bigr) \cdot \dvec \bigl( \tfrac{2\alpha}{\pi} \mu_5 \bigr)
    +  \bigl( \dvec D_5 \bigr) \cdot \dvec \bigl( \tfrac{2\alpha}{\pi} \mu_5 \bigr)
     + \tfrac{2\alpha}{\pi} \mu_5\tfrac{2}{T} \bigl( \dvec T \bigr) \cdot \dvec \bigl(D_5 \bigr)
    } 
    \com\nonumber
    \\ 
    \label{eq:cMHD_final_NS}
    \rho_\mathrm{fl} \bigl( \tfrac{\partial}{\partial t} + \uvec \cdot \dvec \bigr) \uvec 
    & = 
    2 \dvec \cdot \bigl( \rho_\mathrm{fl} \nu \Ssf \bigr) 
    - \tfrac{1}{4} \dvec \rho_\mathrm{fl} 
    + \tfrac{1}{3} \uvec \dvec \cdot (\rho_\mathrm{fl} \uvec)
    \\ & \quad 
    + \tfrac{3}{4} \Jvec \times \Bvec 
    - \uvec \bigl[ \uvec \cdot (\Jvec \times \Bvec) + \eta |\Jvec|^2 \bigr] 
    + {\color{red} \eta \bigl( \tfrac{2\alpha}{\pi} \mu_5 \bigr) \Bvec \cdot \Jvec \, \uvec}\nonumber\com\\
    \label{eq:cMHD_final_drhofl_dt}
    \tfrac{\partial}{\partial t} \rho_\mathrm{fl} 
    & = 
    - \tfrac{4}{3} \dvec \cdot (\rho_\mathrm{fl} \uvec)
    + \uvec \cdot (\Jvec\times\Bvec) 
    + \eta |\Jvec|^2 
    - {\color{red} \eta \bigl( \tfrac{2\alpha}{\pi} \mu_5 \bigr) \Bvec \cdot \Jvec}\per
\end{align}
\end{subequations} 
Now one can easily verify that energy, momentum, and chirality satisfy the desired local conservation laws (to leading order in the small $\mu_5/T$ and small $|\uvec|$ perturbative expansions).
Notice that these equations are written in terms of the three variables $\mu_5$, $T$, and $\rho_\mathrm{fl}$, which are not independent. 
Using \eref{eq:rho_def}, we eliminate $\rho_\mathrm{fl}$ in \eqref{eq:cMHD_final_drhofl_dt}, which allows the second and fourth equations to be written equivalently as
\begin{subequations}\label{eq:dmu5dt_and_dTdt}
\begin{align}
    \tfrac{\partial}{\partial t} \bigl(  \tfrac{2\alpha}{\pi} \mu_5 \bigr) 
    & = 
    - \uvec \cdot \dvec \bigl(  \tfrac{2\alpha}{\pi} \mu_5 \bigr)
    - \tfrac{1}{3} \bigl( \tfrac{2\alpha}{\pi} \mu_5 \bigr) \dvec \cdot \uvec 
    + D_5 \nabla^2 \bigl( \tfrac{2\alpha}{\pi} \mu_5 \bigr) 
    \\ & \quad 
    - \lambda \eta \bigl( \tfrac{2\alpha}{\pi} \mu_5 \Bvec -\Jvec \bigr) \cdot \Bvec 
    - \Gamma_5 \bigl( \tfrac{2\alpha}{\pi} \mu_5 \bigr) 
    - \tfrac{15}{\pi^2 g_E T^4} \bigl( \tfrac{2\alpha}{\pi} \mu_5 \bigr) \Jvec \cdot \bigl(\eta \Jvec -\uvec \times \Bvec\bigr)
    \nn & \quad 
    + \tfrac{2}{3} \tfrac{1}{T} \bigl( \tfrac{2\alpha}{\pi} \mu_5 \bigr) \, \uvec \cdot \dvec T
    + D_5 \tfrac{1}{T^2} \bigl( \tfrac{2\alpha}{\pi} \mu_5 \bigr) \nabla^2 \bigl( T^2 \bigr) 
    + 
    D_5 \tfrac{4}{T} \, \dvec T \cdot \dvec \bigl( \tfrac{2\alpha}{\pi} \mu_5 \bigr)
    \nonumber\\
    &\quad+  \bigl( \dvec D_5 \bigr) \cdot \dvec \bigl( \tfrac{2\alpha}{\pi} \mu_5 \bigr)
     + \tfrac{2\alpha}{\pi} \mu_5\tfrac{2}{T} \bigl( \dvec T \bigr) \cdot \dvec \bigl(D_5 \bigr)
    \com\nonumber
    \\ 
    \tfrac{\partial}{\partial t} T 
    & = 
    - \tfrac{T}{3} \dvec \cdot \uvec 
    - \tfrac{4}{3} \uvec \cdot \dvec T
    + \tfrac{15}{2 \pi^2 g_E} \tfrac{1}{T^3} \,\Jvec \cdot \bigl( \eta \Jvec - \uvec \times \Bvec\bigr) -\tfrac{30}{\pi^2 g_E}\tfrac{1}{T^3}\eta \bigl( \tfrac{2\alpha}{\pi} \mu_5 \bigr)  \Bvec \cdot\Jvec
    \per 
\end{align}
\end{subequations}
These equations in \eqref{eq:dmu5dt_and_dTdt} combined with Maxwell's equation in \eqref{eq:cMHD_final_Maxwell} and the Navier-Stokes equation in \eqref{eq:cMHD_final_NS} constitute our fully consistent, nonrelativistic \cMHD{} equations.

\section{Numerical examples}
\label{sec:examples}

In this section, we use two examples to study how energy is exchanged between the fluid and the electromagnetic field while the chiral plasma instability develops.
First, we consider an artificial monochromatic magnetic field to study the qualitative behavior of the system.
Then, we solve the \cMHD{} equations of motion for the full spectrum.
For both examples, we simplify the calculation by assuming a homogeneous plasma at rest, 
\begin{equation}
    \label{eq:homog_approx}
    T(\xvec,t) = \overline{T}(t) 
    \com \quad 
	\mu_5(\xvec,t) = \overline{\mu}_5(t) 
	\com \quad \text{and} \quad 
    \uvec(\xvec,t) = \overline{\uvec}(t)={\bm 0}
    \per
\end{equation}
Here, and in other derivative quantities, we use a bar to indicate the homogeneous quantities.

\subsection{Monochromatic spectrum}
\label{sub:mono_spectrum}

Since the plasma is electrically neutral, both the electric and magnetic fields are divergence-less, so both fields can be expanded in terms of two circular polarization modes $\evec_\pm$.
In this section, we assume that both fields are monochromatic and occupy the same circular polarization mode, which corresponds to setting  
\bes{\label{eq:monochromatic}
    \Evec(\xvec,t) = \mathrm{Re}\bigl[\sqrt{2} E_{\star,+}(t) \, \ee^{\ii \kvec_\star \cdot \xvec} \, \evec_+ \bigr]
    \qquad \text{and} \qquad  
    \Bvec(\xvec,t) = \mathrm{Re}\bigl[\sqrt{2} B_{\star,+}(t) \, \ee^{\ii \kvec_\star \cdot \xvec} \, \evec_+ \bigr] 
    \com 
}
where $\kvec_\star$ is the wave-number vector of the single nonzero mode, $\evec_+ = \tfrac{1}{\sqrt{2}}({\bm n} + \ii \hat\kvec_\star\times {\bm n})$ with ${\bm n}$ being any real unit vector such that $\evec_+\cdot\kvec_\star=0$, and $E_{\star,+}$ and $\,B_{\star,+}$ are the time-dependent amplitudes of the electric and magnetic fields.
\Eref{eq:monochromatic} simply corresponds to plane waves with a wave-number vector of $\kvec_\star$.
Although the fields vary with $\xvec$, it is important that their dot products are homogeneous: ${\Evec\cdot\Evec = E_{\star,+}^2}$, $\Evec\cdot\Bvec = E_{\star,+}B_{\star,+}$, and $\Bvec\cdot\Bvec = B_{\star,+}^2$.  
Consequently, the electromagnetic energy density is also homogeneous, $\overline{\rho}_\mathrm{em}=\tfrac{1}{2}(E_{\star,+}^2+B_{\star,+}^2)$. 
\Eref{eq:monochromatic} is thus compatible with the homogeneity assumptions.

With the approximations above, in the monochromatic case, the \cMHD{} equations given in \eref{eq:cMHD_corrected} reduce to:
\begin{subequations}
\label{eq:mode_eqns_monochromatic}
\begin{empheq}[box=\fbox]{align} 
    \tfrac{\dd}{\dd t} \overline{T} 
    & = 
    \biggl( \frac{15}{2 \pi^2 g_E} \biggr) \Bigl( \overline{\sigma} \, \overline{T}^{-3} E_{\star,+}^2 \Bigr) 
    - \biggl( \frac{30 \alpha}{\pi^3 g_E} \biggr) \Bigl( \overline{T}^{-3} \, \overline{\mu}_5 E_{\star,+} B_{\star,+} \Bigr) 
    + \mathcal{O}(\overline{\mu}_5^2) \\
    \tfrac{\dd}{\dd t} \overline{\mu}_5 
    & = 
    \biggl( \frac{6\alpha}{\pi} \biggr) \Bigl( \overline{T}^{-2} E_{\star,+} B_{\star,+} \Bigr) 
    - \biggl( \frac{15}{\pi^2 g_E} \biggr) \Bigl( \overline{\sigma} \, \overline{T}^{-4} \, \overline{\mu}_5 E_{\star,+}^2 \Bigr) 
    + \mathcal{O}(\overline{\mu}_5^2) 
    \label{eq:hom_mu_evo} \\
    \tfrac{\dd}{\dd t} B_{\star,+} & =- |\kvec_\star| \, E_{\star,+} \label{eq:hom_B_evo} \\
    \tfrac{\dd}{\dd t} E_{\star,+} & = \bigl( |\kvec_\star| - \tfrac{2\alpha}{\pi} \overline{\mu}_5 \bigr) B_{\star,+} - \overline{\sigma} E_{\star,+} 
    \label{eq:hom_E_evo}
    \per
\end{empheq}
\end{subequations}
Note that the evolution equation for the bulk velocity is identically vanishing: 
the right-hand side of \eref{eq:cMHD_final_NS} is zero in the homogeneous and dissipationless case if $\uvec=0$ and the electromagnetic fields are given by \eref{eq:monochromatic}, in particular $\Jvec\times\Bvec\propto \Evec\times\Bvec={\bm 0}$.

Throughout the evolution, we may define two constants of motion:
one related to the energy-momentum continuity equation, and another one related to the anomaly equation.
As discussed in the previous section, the \cMHD{} equations, as presented here, exactly satisfy the energy-momentum continuity equation, thus with the approximations of \eref{eq:homog_approx} the total energy density is conserved as
\bes{\label{eq:drhobartot}
    \tfrac{\dd}{\dd t} \big(\overline{\rho}_T + \overline{\rho}_5 + \overline{\rho}_\mathrm{em}\big) = 0
    \per
}
The conserved quantity in the anomaly equation is related to the magnetic helicity density, as introduced in \eref{eq:dn5dt_and_dhdt}.
Since the monochromatic fields in \eref{eq:monochromatic} are transverse, in the Weyl gauge, we find the vector potential to be
\bes{
    \Avec(\xvec,t) = 
    \tfrac{1}{|\kvec_\star|}\Bvec(\xvec,t)
    \per
}
Consequently, $\kvec_\star\cdot\Avec=\kvec_\star\cdot\Evec=0$ and we find from \eref{eq:dn5dt_and_dhdt}:
\bes{
    \label{eq:helicity_conservation_equation}
    \tfrac{\partial}{\partial t} \bigl( \overline{n}_5 + 4 \tfrac{\alpha}{4\pi} \overline{h} \bigr) = 0
    \com
}
if we assume that $\Gamma_5=0$.
Otherwise, the quantity on the left-hand side is approximately conserved on times scales that are short compared to the chiral erasure time $\Gamma_5^{-1}$.

In order to discuss the chiral plasma instability, it is useful to define 
\begin{equation}
    \label{eq:k_CPI_and_t_CPI_defs}
    k_\CPI{} = \frac{2\alpha}{\pi} \, \overline{\mu}_5 
    \qquad \text{and} \qquad 
    t_\CPI{} = \frac{4 \, \overline\sigma}{k_\CPI{}^2} = \frac{\pi^2}{\alpha^2} \frac{\overline{\sigma}}{\overline{\mu}_5^2} 
    \per
\end{equation}
We will see that $l_\CPI{}(t) = 2\pi/k_\CPI{}(t)$ sets the instability length scale and $t_\CPI{}(t)$ sets the instability growth time scale.

\subsubsection{Analytical solution}
\label{sec:analytic_sol}

In the homogeneous and monochromatic case, the equations of motion have a simple analytic solution when certain, physically motivated assumptions are made.
First, we assume that the conductivity $\overline{\sigma} \gg \overline{T}$ is large and thus the use of the ideal MHD limit $\tfrac{\partial}{\partial t}\Evec=0$ is justified.
For our equations of motion, this allows us to neglect the $\tfrac{\dd}{\dd t} E_{\star,+}$ term in \pref{eq:hom_E_evo}, which is solved to find $E_{\star,+} \approx \overline{\eta} \bigl( |\kvec_\star| - \tfrac{2\alpha}{\pi} \overline{\mu}_5 \bigr) B_{\star,+}$.  
Second, we assume that $\overline{T}(t) \approx T_i$ remains constant throughout the evolution.
This is justified by the chiral asymmetry being very small, $\overline\mu_5 / T_i \ll 1$, implying only a small amount of heating during chiral erasure.
Subject to these approximations, \eref{eq:mode_eqns_monochromatic} implies 
\bes{\label{eq:monochromatic_ideal_approx}
    \tfrac{\dd}{\dd t} B_{\star,+}(t) 
    = - |\kvec_\star| \, \overline{\eta} \, \bigl( |\kvec_\star| - k_\CPI{}(t) \bigr) \, B_{\star,+}(t) 
    \com
}
where $k_\CPI{}(t)$ was defined in \eref{eq:k_CPI_and_t_CPI_defs}.  
For the initial condition $B_{\star,+}(t_i)=B_{\star,+,i}$, the solution of \eref{eq:monochromatic_ideal_approx} is 
\begin{subequations}
\label{eq:B_mono_sol}
\begin{align}
    B_{\star,+}(t) & = \exp\bigg[- \frac{4 |\kvec_\star| \big(|\kvec_\star| - \overline{k}_\CPI{}(t) \big)}{k_\CPI{}(t_i)^2} \frac{t-t_i}{t_\CPI{}(t_i)} \bigg] \, B_{\star,+,i} \com \\ 
    \overline{k}_\CPI{}(t) & = \frac{1}{t-t_i} \int_{t_i}^t \! \dd  t^\prime \, k_\CPI{}(t^\prime) 
    \per
\end{align}    
\end{subequations}
If $\overline{\mu}_5(t)$ were constant, then \eref{eq:B_mono_sol} would reduce to \eref{eq:gen_ana_B_Field}. 
This solution shows the aforementioned exponential growth of the magnetic field for $|\kvec_\star|<\overline k_\CPI{}(t)$, however, in order to find the full solution we also require the explicit time evolution of the chiral chemical potential.

We can infer the evolution of the chiral chemical potential using \eref{eq:dn5dt_and_dhdt}. 
If chirality-violating scatterings are out of equilibrium in the plasma, then $\overline{n}_5(t) + 4 \tfrac{\alpha}{4\pi} \overline{h}(t)$ is approximately conserved. 
Equating this quantity at time $t_i$ and time $t \neq t_i$ implies 
\begin{equation}
    \label{eq:kCPI_equation_helicity_conservation}
    \frac{1}{24} \, \frac{4\pi}{\alpha} \, k_{\CPI{}}(t_i) \, T_i^2 
    + 4 \frac{\alpha}{4\pi} \, \frac{B_{\star,+}^2(t_i)}{|\kvec_\star|} 
    \approx 
    \frac{1}{24} \, \frac{4\pi}{\alpha} \, k_\CPI{}(t) \, T_i^2 
    + 4 \frac{\alpha}{4\pi} \, \frac{B_{\star,+}^2(t)}{|\kvec_\star|} 
    \com 
\end{equation}
where $B_{\star,+}(t)$ is given by \eref{eq:B_mono_sol}. 
This relation is an integral equation for $k_\CPI{}(t)$ that can be solved analytically.
To simplify expressions, we introduce the following dimensionless quantities:
\begin{equation}
    \label{eq:dimensionless_defs}
    \kappa_\star =\frac {|\kvec_\star|}{k_\CPI{}(t_i)} \com \qquad \tau = \frac{t}{t_\CPI{}(t_i)} \com \qquad \xi= \frac{B_{\star,+,i}^2}{2} \left[\frac{\mu_i^2 T_i^2}{3}\right]^{-1} \approx \frac{3\overline\rho_\mathrm{em}(t_i)}{2\overline\rho_{\rm 5}(t_i)}\,.
\end{equation}
Then, the solution to \eref{eq:kCPI_equation_helicity_conservation} may be written as
\begin{equation}\label{eq:kappa5_sol}    
    \frac{k_\CPI{}(t)}{k_\CPI{}(0)}
    = \kappa_\star +  
    \dfrac{(1 -\kappa_\star) \bigl[ 1 + \kappa_\star (1-\kappa_\star) \xi^{-1} \bigr]}{\kappa_\star (1 - \kappa_\star) \xi^{-1} + \exp\big[8 \big( 1 + \kappa_\star (1 - \kappa_\star) \xi^{-1} \bigr) \xi \tau \big]} 
    \com 
\end{equation}
where we have also set $t_i = 0$.
\Erefs{eq:B_mono_sol}{eq:kappa5_sol} fully determine the evolution of the approximate system.

\begin{figure}[t]
    \centering
    \includegraphics[width=0.95\columnwidth]{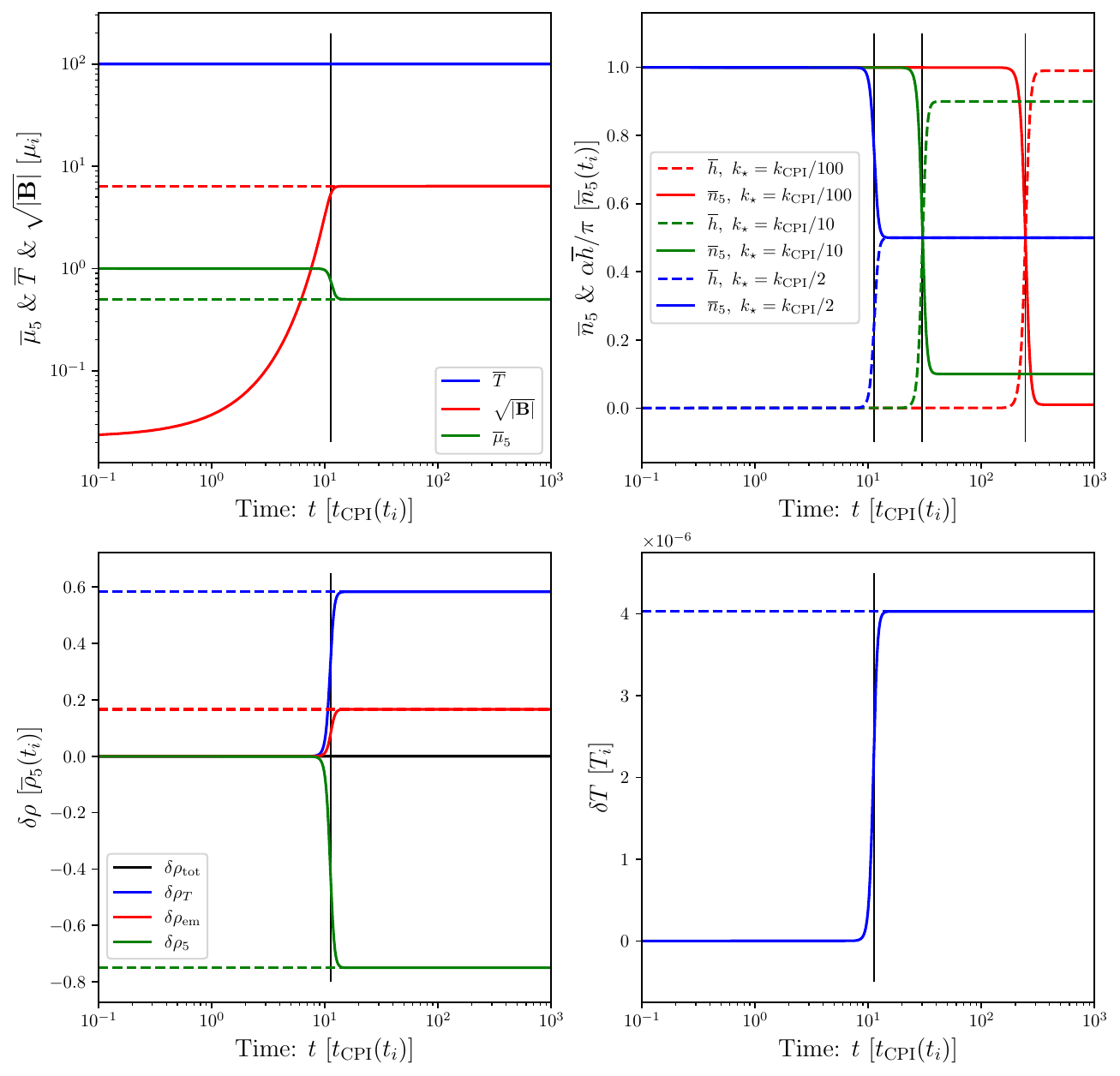}
    \caption{\label{fig:mono_evolution}
    Evolution of various quantities while the \CPI{} develops, as function of time $t$ in units of the instability time scale $t_\CPI{}(t_i)$.  
    \textit{Top-left:}  Evolution of the square root of the magnetic field strength $\sqrt{B_{\star,+}(t)}$, the chiral chemical potential $\overline{\mu}_5(t)$, and the temperature $\overline{T}$ in units of the initial chiral chemical potential $\mu_i$.  
    \textit{Top-right:}  Evolution of the chiral number density $\overline{n}_5(t)$ and the magnetic helicity density $\overline{h}(t)$ in units of $\overline{n}_5(t_i)$ for several values of the wavenumber $|\kvec_\star|$.
    \textit{Bottom-left:}  Evolution of the thermal energy density $\overline{\rho}_T(t)$, the electromagnetic energy density $\overline{\rho}_\mathrm{em}(t)$, the chiral energy density $\overline{\rho}_5(t)$, and the total energy density $\overline{\rho}_\mathrm{tot}(t)$. 
    We show the change of each energy density relative to its initial value, i.e. $\delta \rho = \overline{\rho}(t) - \overline{\rho}(t_i)$, and normalized to the initial chiral energy.    
    \textit{Bottom-right:}  Evolution of the relative change in plasma temperature $\dT(t) = \overline{T}(t) - T_i$. 
    Where not mentioned, we fixed the wave-number to $|\kvec_\star|=k_\CPI{}/2$.
    The vertical black lines are the time at which the CPI is developing the fastest. They are calculated using~\eref{eq:instability_scale_expression_general}.
    }
\end{figure}

\subsubsection{Initial conditions}
\label{sub:mono_initial}

In order to study how the various energy components evolve while the CPI develops, we solve the equations of motion \pref{eq:mode_eqns_monochromatic} numerically.  
We take the initial conditions at $t=t_i$ to be 
\bsa{eq:mono_initial}{
    \overline{T}(t_i) & = 8 \times 10^4 \GeV \equiv T_i \\ 
    \overline{\mu}_5(t_i) & = 10^{-2} \, T_i \equiv \mu_i \\ 
    B_{\star,+}(t_i) & = + 5 \times 10^{-8} \, T_i^2 \equiv B_{\star,+,i} \\ 
    E_{\star,+}(t_i) & = -5 \times 10^{-13} \, T_i^2 \equiv E_{\star,+,i} 
    \per
}
We also set $g_E =2+4\cdot\tfrac{7}{8} = \tfrac{11}{2}$, corresponding to a QED plasma of photons with relativistic electrons and positrons, and we take the electric conductivity to be $\overline{\sigma} = 100 \overline{T}$ \cite{Arnold:2000dr}. 
The initial plasma temperature $T_i$ sets the overall scale. 
For $T_i \gtrsim 80 \TeV$ the chirality-violating interactions are out of equilibrium, and it is a good approximation to set $\overline{\Gamma}_5 = 0$.   
The initial field values $B_{\star,+,i}$ and $E_{\star,+,i}$ are chosen to be small but nonzero, so that they can seed the instability. 
We have defined our magnetic field such that if we have a thermal distribution, i.e. $B_i = \sqrt{\frac{2}{\pi^2}\frac{k^4}{\exp(k/T_i)-1}}$, similar to how we will do in section~\ref{sub:chrom_spectrum}, our initial magnetic field has the same amplitude as for the wavenumber $k=k_\CPI{}/2$. The electric field was chosen to be several orders of magnitude less than the magnetic field as to ensure that Ohm's law does not deplete the energy in the fields before the chiral instability is able to grow the magnetic field. This is physically motivated as in ideal plasma the electric field is suppressed. The electric field and the magnetic field were chosen to have opposite signs as this causes the chiral plasma instability to develop.
Our numerical solutions are summarized in the four panels of \fref{fig:mono_evolution}.

\subsubsection*{Top-left panel}

The top-left panel of \fref{fig:mono_evolution} shows the evolution of the square root of the magnetic field strength $\sqrt{B_{\star,+}(t)}$, the chiral chemical potential $\overline{\mu}_5(t)$, and the temperature $\overline{T}_i$ for $|\kvec_\star| = k_\CPI{}(t_i)/2$.
In the figure, we observe that the magnetic field and chiral chemical potential both saturate to constant values at late times.
This behavior can be understood from the approximate analytical solutions in \erefs{eq:B_mono_sol}{eq:kappa5_sol}.  
First, note that the system of equations \pref{eq:mode_eqns_monochromatic} has an infinite family of non-trivial fixed points, \ie{}, a set of values for the dynamical variables at which all of their time derivatives vanish, parameterized by fixed values of $E_{\star,+} \to 0$ and $\overline\mu_5 \to \tfrac{\pi}{2\alpha} |\kvec_\star|$, and arbitrary values of $B_{\star,+}$ and $\overline{T}$.
The fixed points should also be reflected in the analytic solution, which is easy to verify using \eref{eq:kappa5_sol}:
\begin{equation} \label{eq:late_time_wavenumber}
    \lim_{t\to\infty} k_\CPI{}(t) = \tfrac{2\alpha}{\pi}\lim_{t\to\infty} \overline{\mu}_5(t) = |\kvec_\star| 
    \per 
\end{equation}
Then the late-time limit of our analytical approximation for the magnetic field \eref{eq:B_mono_sol} with the dimensionless quantities defined in \eref{eq:dimensionless_defs} is 
\begin{equation}\label{eq:late_analytic_BField}
    \lim_{t\to\infty}B_{\star,+}(t) = \sqrt{1 + \xi^{-1} \kappa_\star (1-\kappa_\star) } \, B_{\star,+,i} \approx 
    \mu_i T_i\sqrt{\tfrac{2}{3}\kappa_\star(1-\kappa_\star)}
    \per
\end{equation}
For most physically relevant scenarios $B_{\star,+,i}^2 \ll \mu_i^2 T_i^2$ (\ie{}, $\xi\ll1$) and consequently the late-time magnetic field strength is largely independent of its initial value $B_{\star,+,i}$.  
In the top-left panel of \fref{fig:mono_evolution}, red and green dashed lines show the analytic late-time limits of $B_{\star,+}$ and $\overline\mu_5$, respectively.

From \eref{eq:kappa5_sol}, we can also analytically infer the timescale for the instability to fully develop.
As the development of the instability is exponential, the change in the chemical potential is highly peaked around a time $t_{\rm inst.}$.
We define this timescale $t_{\rm inst.}$ as the time when the change in $k_\CPI{}(t)$ is maximal, \ie{},
\begin{equation}
    \frac{\dd^2 k_\CPI{}(t)}{\dd t^2}\Big|_{t=t_{\rm inst.}} = 0 
    \qquad \text{and} \qquad 
    \frac{\dd k_\CPI{}(t)}{\dd t}\Big|_{t=t_{\rm inst.}} < 0 
    \per
\end{equation}
For $0\leq|\kvec_\star|<k_\CPI{}(t_i)$ (where the instability is relevant) the timescale is found to be
\begin{equation}
    \label{eq:instability_scale_expression_general}
    t_{\rm inst.} = \frac{t_\CPI{}(t_i)}{8\big[\kappa_\star(\kappa_\star-1)-\xi\big]}\ln\left[\frac{\xi}{\kappa_\star(1-\kappa_\star)}\right]
    \per
\end{equation}
Notice that the derived timescale depends on the initial magnetic field strength through the parameter $\xi \propto B_{*,+,i}^2$.
A larger initial magnetic field leads to a faster development of the instability, whereas a smaller initial field leads to a slower instability.
Note that for $\xi > \kappa_\star(1 - \kappa_\star)$ -- that is, when the initial magnetic field energy is not negligible compared to the chiral energy -- the solution to \eref{eq:instability_scale_expression_general} becomes negative. 
This implies that physically relevant initial conditions satisfy $\xi < \kappa_\star(1 - \kappa_\star)$; otherwise, the system would already be highly unstable at the outset, which is difficult to imagine arising from a physical system.
As in the top-left panel of \fref{fig:mono_evolution}, we take $\xi \ll 1$ and $|\kvec_\star| = k_\CPI{}(t_i) / 2$, then we find
\begin{equation}
\label{eq:instability_scale_expression_max}
    t_{\rm inst.}|_{\kappa_\star=1/2} \approx -\frac{1}{2}\ln (4\xi) \, t_\CPI{}(t_i) 
    \per
\end{equation}
For the initial conditions in \eref{eq:mono_initial}, this quantity evaluates to $t_{\rm inst.}|_{\kappa_\star=1/2} \approx 11.3 \, t_\CPI{}(t_i)$. 
By inspecting the various panels of \fref{fig:mono_evolution}, we observe that the instability saturation time agrees well with the analytical expression in \eref{eq:instability_scale_expression_max}, which is drawn as the vertical black lines.

\subsubsection*{Top-right panel}

The top-right panel of \fref{fig:mono_evolution} shows the evolution of chiral number density $\overline n_5$ and the magnetic helicity density $\overline h$ for various values of the wave-number $|\kvec_\star|$.
From \eref{eq:helicity_conservation_equation}, we have the conservation law
\begin{equation}
    \overline n_5 + \tfrac{\alpha}{\pi}\overline h = {\rm constant}\com
\end{equation}
which is reflected also in the numerical solution.
In the figure, we see that the chiral number density decreases in time and the magnetic helicity density increases by the same amount, irrespectively of the wave-number.
The wave-number, however, does affect the final saturation densities.
The magnetic helicity density in the monochromatic case can simply be calculated at late times from \eref{eq:late_analytic_BField}:
\begin{equation}
    \lim_{t\to\infty}\overline h(t) = |\kvec_\star|^{-1} \lim_{t\to\infty} B_{\star,+}^2(t) \approx \left(1-\frac{|\kvec_\star|}{k_\CPI{}(t_i)}\right)\overline n_5(t_i)\per
\end{equation}
Similarly, from \eref{eq:late_time_wavenumber} we can obtain the late-time chiral number density with the approximation $\overline{T}(t)\approx T_i$:
\begin{equation}
    \lim_{t\to\infty}\overline n_5(t) \approx \tfrac{1}{3}\lim_{t\to\infty}\overline \mu_5(t) \overline{T}^2(t) \approx \frac{|\kvec_\star|}{k_\CPI{}(t_i)} \overline n_5(t_i) \per  
\end{equation}
As we decrease the wave-number of the monochromatic spectrum, the chiral number density is depleted further, and correspondingly, the magnetic helicity density is amplified.
At $|\kvec_\star|=k_\CPI{}/2$, the final values for the densities are equal, which is also confirmed in the numerical result in \fref{fig:mono_evolution}.

\subsubsection*{Bottom-left panel}

The bottom-left panel of \fref{fig:mono_evolution} shows the evolution of the various energy components of the chiral plasma.     
In this figure, we are plotting the change  in the energy density ($\delta \rho(t) \equiv \overline{\rho}(t) - \overline{\rho}(t_i)$) relative to its initial value at time $t_i=0$ in units of the initial chiral energy density $\overline{\rho}_5(t_i) = \tfrac{1}{2} \mu_i^2 T_i^2$.  
We observe that when the \CPI{} develops, the thermal energy increases, the electromagnetic energy increases, the chiral energy decreases, and the total energy remains constant.  
After the instability saturates, all of the energy components remain constant. 
This can be taken as evidence that our equations of motion describe the complete energy balance of the system.  
More precisely, we can see that the chiral instability provides the energy for the growing magnetic field, but we can also see that most of the energy from the chiral instability goes into heating the plasma, \ie{} $\delta \rho_T$ is larger than $\delta \rho_\mathrm{em}$.

Using the analytical solutions discussed in the previous sections, it is straightforward to derive expressions for the energy changes.  
In particular, we use \erefs{eq:late_time_wavenumber}{eq:late_analytic_BField}, which give values of $B_{\star,+}$ and $\overline{\mu}_5$ after the instability saturates.  
Using the analytic solutions for $B_{\star,+}$ and $\overline{\mu}_5$ to calculate the various energies gives 
\bsa{eq:energy_changes}{
    \lim_{t\to\infty} \frac{\delta\rho_5(t)}{\overline{\rho}_5(t_i)} & = - \biggl( 1 + \frac{|\kvec_\star|}{k_\CPI{}(t_i)} \biggr) \biggl( 1 - \frac{|\kvec_\star|}{k_\CPI{}(t_i)} \biggr) \com \\
    \lim_{t\to\infty}\frac{\delta\rho_\mathrm{em}(t)}{\overline{\rho}_5(t_i)} & = \frac{2}{3} \biggl( \frac{|\kvec_\star|}{k_\CPI{}(t_i)} \biggr) \biggl( 1 - \frac{|\kvec_\star|}{k_\CPI{}(t_i)} \biggr) \com \\
    \lim_{t\to\infty}\frac{\delta\rho_{T}(t)}{\overline{\rho}_5(t_i)} & = \biggl( 1 + \frac{1}{3}  \frac{|\kvec_\star|}{k_\CPI{}(t_i)} \biggr) \biggl( 1 - \frac{|\kvec_\star|}{k_\CPI{}(t_i)} \biggr) 
    \per
}
The last line is obtained by imposing energy conservation.
Note, that $\lim_{t\to\infty}\delta\rho_{T}(t) > \lim_{t\to\infty} \delta \rho_\mathrm{em}(t)$ for all wavenumbers in $0 \leq |\kvec_\star| < k_\CPI{}(t_i)$, meaning that more energy is absorbed in the background than what is used for magnetic field amplification.
By inspecting the bottom-left panel of \fref{fig:mono_evolution}, we observe that the saturated energy densities agree well with the analytical expressions in \eref{eq:energy_changes}, which are drawn as the horizontal dashed lines. 

\subsubsection*{Bottom-right panel}

The bottom-right panel of \fref{fig:mono_evolution} shows the evolution of the plasma temperature relative to its initial value, $\dT(t) = \overline{T}(t) - T_i$.
The temperature increases while the CPI develops.
In particular, similar to the other parameters, the temperature exponentially increases at about $t\approx10 t_\mathrm{CPI}$ and then saturates.
This panel shows the same $T$ evolution as the top-left panel of \fref{fig:mono_evolution}, which shows that the temperature change is small in comparison to $T_i$.

Using the analytical solutions discussed in the previous sections, it is straightforward to derive an expression for the temperature change.  
It follows immediately from the expression for $\delta \rho_T$ in \eref{eq:energy_changes}, which gives 
\bes{\label{eq:dT_mono}
    \lim_{t \to \infty} \dT(t) & = \frac{15 \mu_i^2}{4\pi^2 g_E T_i} \biggl( 1 + \frac{1}{3} \frac{|\kvec_\star|}{k_\CPI{}(t_i)} \biggr) \biggl( 1 - \frac{|\kvec_\star|}{k_\CPI{}(t_i)} \biggr)  
    \per
}
This quantity is positive for all $0 \leq |\kvec_\star| < k_\CPI{}(t_i)$, indicating that the plasma is only heated and never cooled.  
Parametrically, the temperature increases by $\dT \sim \mu_i^2 / g_E T_i$.  
By inspecting the bottom-right panel of \fref{fig:mono_evolution}, we observe that the saturated temperature change agrees well with the analytical expression in \eref{eq:dT_mono}, which is drawn as the blue-dashed line.

\subsection{Non-monochromatic spectrum}
\label{sub:chrom_spectrum}

In order to provide a second illustration of \CPI{}-induced heating, in this section we study a somewhat more realistic modeling of the electromagnetic field.  
Whereas in \eref{eq:monochromatic} we restricted the field to a single Fourier mode $\kvec = \kvec_\star$, in this section we relax that restriction and allow for a non-monochromatic spectrum.  
In general, for a charge-neutral plasma, the electric and magnetic fields may be decomposed onto Fourier modes labeled by wavevector $\kvec$, and then further decomposed onto a basis of right- and left-circular polarization modes, labeled by $\pm$.  
This is written as 
\bes{\label{eq:Fourier}
    \Xvec(\xvec,t) 
    = \int \! \! \frac{\dd^3 \kvec}{(2\pi)^3} \, \tilde{\Xvec}(\kvec,t) \, \ee^{\ii \kvec \cdot \xvec} 
    \qquad \text{and} \qquad 
    \tilde{\Xvec}(\kvec,t) = \tilde{X}_{\kvec,+}(t) \, \evec_{\kvec,+} + \tilde{X}_{\kvec,-}(t) \, \evec_{\kvec,-} 
    \com
}
where $\Xvec$ stands for either $\Evec$ or $\Bvec$, $\tilde{\Xvec}(\kvec,t)$ is the Fourier transform of $\Xvec(\xvec,t)$, and $\tilde{X}_{\kvec,\pm} = \evec_{\kvec,\pm}^\ast \cdot \tilde{\Xvec}(\kvec,t)$, with $\evec_{\kvec,\pm}$ defined as below \eref{eq:monochromatic}. 
Since $\Xvec(\xvec,t)^\ast = \Xvec(\xvec,t)$, 
$\tilde{\Xvec}(\kvec,t)^\ast = \tilde{\Xvec}(-\kvec,t)$.  
Neither $\tilde{\Evec}$ nor $\tilde{\Bvec}$ contain a longitudinal polarization mode; this component is identically zero for the magnetic field because $\dvec \cdot \Bvec = 0$, and it is zero for the electric field because we assume charge neutrality $\dvec \cdot \Evec = Q = 0$. 
We model the electromagnetic fields as stochastic variables drawn from a time-dependent ensemble that is statistically homogeneous and isotropic.  
This lets us write the equal-time two-point correlation function as 
\bes{
    \langle X_i(\xvec,t) \, Y_j(\yvec,t) \rangle = C_{XY,ij}(|\xvec-\yvec|,t)
    \com
}
which only depends on the spatial coordinates, $\xvec$ and $\yvec$, through the function $|\xvec-\yvec|$.  
The Fourier representation of this relation is written as 
\ba{
    C_{XY,ij}(|\xvec-\yvec|,t) & = \int \! \! \frac{\dd^3 \kvec}{(2\pi)^3} \int \! \! \frac{\dd^3 \qvec}{(2\pi)^3} \, \langle \tilde{X}_i(\kvec,t) \, \tilde{Y}_j(\qvec,t)^\ast \rangle \, \ee^{\ii \kvec \cdot \xvec - \ii \qvec \cdot \yvec}, \\ 
    \langle \tilde{X}_i(\kvec,t) \, \tilde{Y}_j(\qvec,t)^\ast \rangle & = (2\pi)^3 \delta^{(3)}(\kvec - \qvec) \, \Biggl[ \biggl( \delta_{ij} - \frac{k_i k_j}{|\kvec|^2} \biggr) \, \mathcal{S}_{XY}(|\kvec|,t) - \ii \epsilon_{ijk} \frac{k_k}{|\kvec|} \, \mathcal{A}_{XY}(|\kvec|,t) \Biggr] 
    \com \nonumber
}
where $\mathcal{S}_{XY}(k,t)$ and $\mathcal{A}_{XY}(k,t)$ are time-dependent power spectra.  
With these definitions $\mathcal{A}_{XX}(k,t) \leq \mathcal{S}_{XX}(k,t)$ for $X = E$ or $B$.
Note that 
\bes{
    \langle \Xvec(\xvec,t) \cdot \Yvec(\yvec,t) \rangle 
    = 2 \int_0^\infty \! \frac{\dd k}{k} \frac{k^3}{2 \pi^2} \, \mathcal{S}_{XY}(k,t) \, \mathrm{sinc}\big( k |\xvec - \yvec| \bigr)
    \com
}
where $\mathrm{sinc}(x) = \sin(x)/x$ for $x \neq 0$ and $\mathrm{sinc}(0) = 1$.  
This makes it possible to define the characteristic electric and magnetic field amplitudes on the length scale 
$l= 2\pi/k$ as 
\bes{
    E_k(t) \equiv \sqrt{ \frac{k^3}{2 \pi^2} \, \mathcal{S}_{EE}(k,t) } 
    \qquad \text{and} \qquad 
    B_k(t) \equiv \sqrt{ \frac{k^3}{2 \pi^2} \, \mathcal{S}_{BB}(k,t) } 
    \per
}
In the Weyl gauge (\ie{}, $V(\xvec,t) = 0$), the magnetic vector potential $\Avec(\xvec,t)$ admits a Fourier decomposition of the form shown in \eref{eq:Fourier} where $\tilde{A}_{\kvec,\pm} = \pm \tilde{B}_{\kvec,\pm} / |\kvec|$.  
It follows that the magnetic helicity density $h(\xvec,t) = \Avec \cdot \Bvec$ has an ensemble average that is calculated as
\bes{\label{eq:chrom_helicity}
    \overline{h}(t) 
    = \langle \Avec(\xvec,t) \cdot \Bvec(\xvec,t) \rangle 
    = \int_0^\infty \frac{\dd k}{k} \frac{k^2}{\pi^2} \, \mathcal{A}_{BB}(k,t) 
    \com
}
which is independent of $\xvec$ for this statistically homogeneous ensemble.  

To study the evolution of this system, we adopt the same assumptions that were enumerated in \sref{sub:mono_initial}, but we relax the restriction of a monochromatic spectrum as described above.  
In particular, we assume that the temperature and chiral chemical potential are homogeneous, which lets us write $T(\xvec,t) = \overline{T}(t)$ and $\mu_5(\xvec,t) = \overline{\mu}_5(t)$.  
However, dot products of the electric and magnetic fields are not homogeneous when we allow for a non-monochromatic spectrum.  
In order to maintain homogeneous equations, we replace dot products with their ensemble averages, which are homogeneous; \eg{}, 
$\Evec \cdot \Evec \to \langle \Evec \cdot \Evec \rangle$ and $\Evec \cdot \Bvec \to \langle \Evec \cdot \Bvec \rangle$. 
Doing so leads to a new system of equations of motion: 
\small
\begin{subequations}\label{eq:eqns_of_motion_chromatic}
\begin{empheq}[box=\fbox]{align}
    \tfrac{\dd}{\dd t} \overline{T} 
    & = 
    -\frac{120}{\pi^2 g_E} \frac{\alpha}{4\pi} \frac{\overline{\mu}_5}{\overline{T}^3} \langle \Evec \cdot \Bvec \rangle 
    + \frac{15}{2 \pi^2 g_E} \frac{\overline{\sigma}}{\overline{T}^3} \langle \Evec \cdot \Evec \rangle  \\
    \tfrac{\dd}{\dd t} \overline{\mu}_5 
    & = 
    - \frac{15}{\pi^2 g_E} \frac{\overline{\sigma}\overline{\mu}_5}{\overline{T}^4} \langle \Evec \cdot \Evec \rangle 
    + 24 \frac{\alpha}{4\pi} \frac{1}{\overline{T}^2} \langle \Evec \cdot \Bvec \rangle \\
    \tfrac{\partial}{\partial t} \mathcal{S}_{BB}(|\kvec|,t) 
    & = - 2|\kvec| \, \mathcal{A}_{EB}(|\kvec|,t) \\ 
    \tfrac{\partial}{\partial t} \mathcal{S}_{EB}(|\kvec|,t) 
    & = - |\kvec|    \mathcal{A}_{EE}(|\kvec|,t) + |\kvec|  \mathcal{A}_{BB}(|\kvec|,t)  - \tfrac{2\alpha}{\pi} \overline{\mu}_5\,  \mathcal{S}_{BB}(|\kvec|,t) -\overline{\sigma}  \mathcal{S}_{EB}(|\kvec|,t)  \\  
    \tfrac{\partial}{\partial t} \mathcal{S}_{EE}(|\kvec|,t)\label{eq:SEE}
    & = 2 |\kvec| \, \mathcal{A}_{EB}(|\kvec|,t)  - 16 \tfrac{\alpha}{4\pi} \overline{\mu}_5\,  \mathcal{S}_{EB}(|\kvec|,t) -2\overline{\sigma}  \mathcal{S}_{EE}(|\kvec|,t)  \\ 
    \tfrac{\partial}{\partial t} \mathcal{A}_{BB}(|\kvec|,t) 
    & = -2 |\kvec| \, \mathcal{S}_{EB}(|\kvec|,t)  \\ 
    \tfrac{\partial}{\partial t} \mathcal{A}_{EB}(|\kvec|,t) 
    & = - |\kvec|    \mathcal{S}_{EE}(|\kvec|,t) + |\kvec|  \mathcal{S}_{BB}(|\kvec|,t)  - \tfrac{2\alpha}{\pi} \overline{\mu}_5\,  \mathcal{A}_{BB}(|\kvec|,t) -\overline{\sigma}  \mathcal{A}_{EB}(|\kvec|,t) \\ 
    \tfrac{\partial}{\partial t} \mathcal{A}_{EE}(|\kvec|,t) 
    & =  2|\kvec| \, \mathcal{S}_{EB}(|\kvec|,t)  - 16 \tfrac{\alpha}{4\pi} \overline{\mu}_5\,  \mathcal{A}_{EB}(|\kvec|,t)-2\overline{\sigma}  \mathcal{A}_{EE}(|\kvec|,t)  \label{eq:AEE}
\end{empheq}
\end{subequations}
\normalsize
where 
\ba{\label{eq:exp_XY}
    \langle \Xvec \cdot \Yvec \rangle 
    & = 2 \int_0^\infty \! \frac{\dd k}{k} \frac{k^3}{2 \pi^2} \, \mathcal{S}_{XY}(k,t) 
    \qquad \text{for $\Xvec,\Yvec = \Evec,\Bvec$} 
    \per
}
The assumption of homogeneity is expected to break down when the turbulent fluid motion becomes significant.  
Therefore, these equations can describe the onset of the \CPI{} and the associated energy transfer, which is our primary interest, but they will not accurately capture the instability's saturation and subsequent development.  

\subsubsection{Initial conditions}
\label{sub:chrom_initial}

In order to study how the various energy components evolve while the CPI develops, we solve the equations of motion \eref{eq:eqns_of_motion_chromatic} numerically.  
We take the initial conditions to be 
\bsa{eq:chrom_initial}{
    t_i & = 0 \\ 
    \overline{T}(t_i) & = 8 \times 10^4 \GeV \equiv T_i \\ 
    \overline{\mu}_5(t_i) & = 8 \times 10^2 \GeV \equiv \mu_i \\
    \mathcal{S}_{BB}(k,t_i) & = \frac{2 k}{\ee^{k/T_i}-1} \equiv \mathcal{S}_{BB,i}(k) \\  
    \mathcal{S}_{EE}(k,t_i) & = \frac{2 k}{\ee^{k/T_i}-1} \equiv \mathcal{S}_{EE,i}(k) \\ 
    \mathcal{S}_{EB}(k,t_i) & = 0 \equiv \mathcal{S}_{EB,i}(k) \\ 
    \mathcal{A}_{BB}(k,t_i) & = 0 \equiv \mathcal{A}_{BB,i}(k) \\  
    \mathcal{A}_{EE}(k,t_i) & = 0 \equiv \mathcal{A}_{EE,i}(k) \\
    \mathcal{A}_{EB}(k,t_i) & = 0 \equiv \mathcal{A}_{EB,i}(k) 
    \per
}
The spectra are chosen such that the fields are initially in a thermal spectrum.  
We have also set $\mathcal{S}_{EB,i}(k)=0$ because the thermal ensemble is parity invariant.

\subsubsection{Evolution of homogeneous quantities}
\label{sub:homo_evolution}

\begin{figure}[t]
    \centering
    \includegraphics[width=0.95\columnwidth]{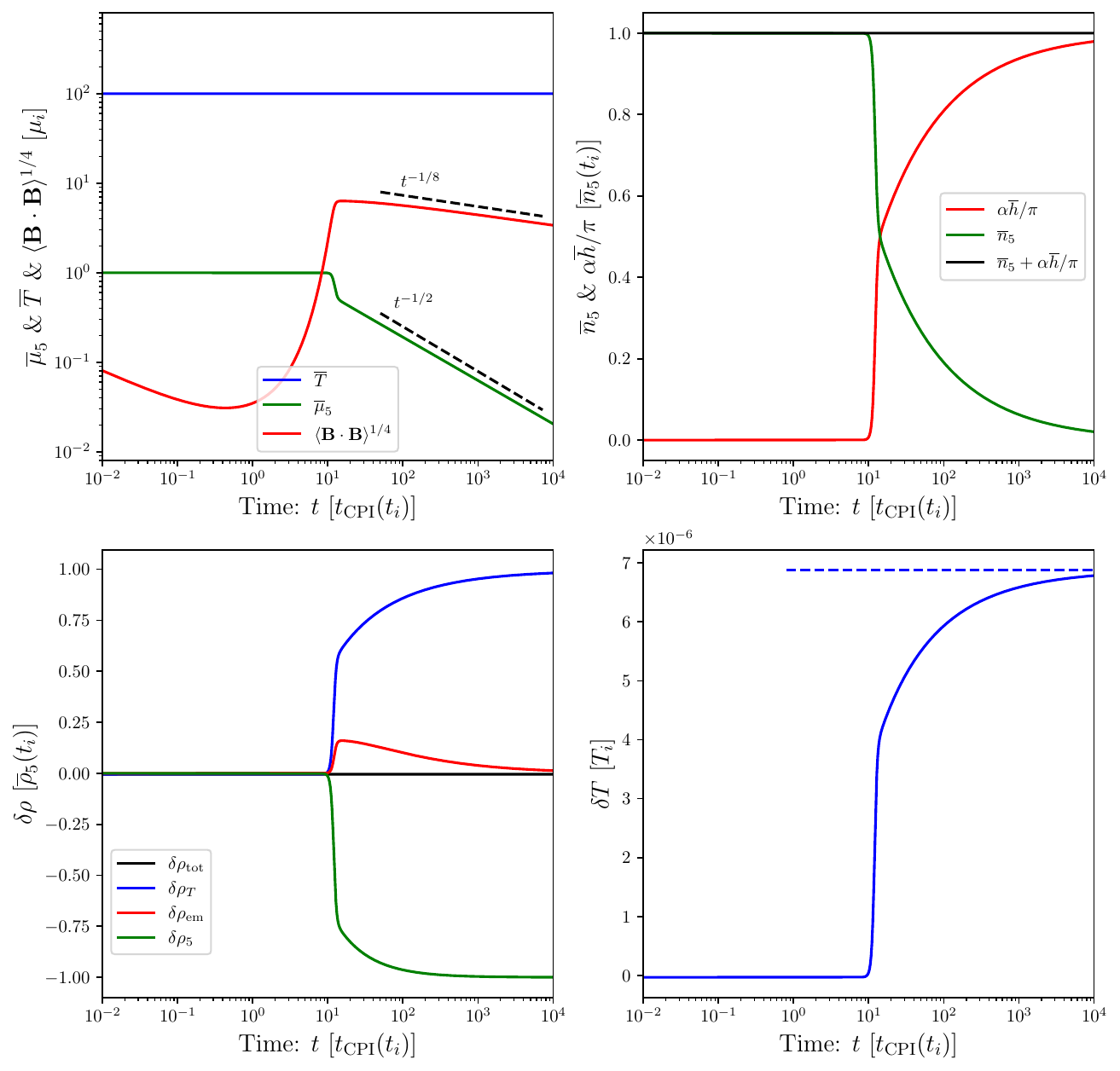}
    \caption{
    \label{fig:chrom_evolution}
    Evolution of various homogeneous quantities during the development of the \CPI{}.  The notation here matches \fref{fig:mono_evolution}.
    }
\end{figure}

\Fref{fig:chrom_evolution} show the evolution of various homogeneous quantities.  
Note that we plot the same quantities in the four panels of \fref{fig:chrom_evolution} (for the non-monochromatic ansatz) as in the four panels of \fref{fig:mono_evolution} (for the monochromatic ansatz), which facilitates a comparison between them. 
The initial evolution of the quantities until the development of the CPI is similar to that of the monochromatic case at $|\kvec_\star|=k_\CPI{}/2$, with a non-trivial late-time evolution due to dissipative effects.
We discuss each of the panels in turn.  

The top-left panel of \fref{fig:chrom_evolution} shows the evolution of the temperature $\overline{T}(t)$, the chiral chemical potential $\overline{\mu}_5(t)$, and the square root of the magnetic field strength $\langle \Bvec \cdot \Bvec \rangle^{1/4}$. 
The temperature appears to remain constant, since the negligible heating cannot be seen on this scale; we show its fractional change in the bottom-right panel, which is discussed below.  
The chiral chemical potential is approximately constant at early times ($t \lesssim 10 t_\CPI{}(t_i)$), abruptly decreases at $t \approx 10 t_\CPI{}(t_i)$ by a factor of $\approx 1/2$, and continues decreasing as approximately $\overline{\mu}_5 \propto t^{-1/2}$.  The $t \approx 10 t_\CPI{}(t_i)$ time scale follows from a similar calculation for the time scale in the monochromatic case in~\eref{eq:instability_scale_expression_max}. 
In \aref{app:chemical_potential_scaling}, we provide an analytical analysis that furnishes an understanding of the late-time scaling.  
In short, the chiral asymmetry depletes at the cost of growing the helical magnetic field, and since the \CPI{} shifts to smaller wavenumbers as the asymmetry is consumed (recall, $k_\CPI{}(t) \propto \overline{\mu}_5(t)$), this depletion does not saturate.  
The magnetic field strength is decreasing at early times ($t \lesssim t_\CPI{}(t_i)$), rises sharply at $t \approx 10 t_\CPI{}(t_i)$, and then continues decreasing as approximately $\langle \Bvec \cdot \Bvec \rangle^{1/4} \propto t^{-1/8}$.
The late-time scaling follows from $\langle \Bvec \cdot \Bvec \rangle \propto \overline{\mu}_5$, which we discuss further in the next section. 
This same scaling was observed by the authors of \rref{Pavlovic:2016gac} when neglecting the fluid velocity as we have done here.  
However, the early time evolution of the magnetic field can be understood as follows.  
The magnetic field decreases at early times because the magnetic field at wave numbers $k > k_\CPI{}(t_i) = \tfrac{2\alpha}{\pi} \mu_i$ is diffused.
By \eref{eq:eqns_of_motion_chromatic} for $k>k_\CPI{}(t_i)$ by approximating $\partial_t\mathcal{A}_{EB}\approx 0$ and $\mathcal{S}_{EE}\approx0$ we can see that,
\bes{
    \frac{\partial}{\partial t}\mathcal{S}_{BB} \approx \frac{-2 |\kvec|^2}{\overline{\sigma}}\mathcal{S}_{BB},
}
which indicates exponential decay of large wavenumbers. As at early time, the $k^3\mathcal{S}_{BB}$ power spectra is peaked at large $k$, this leads to an initially decreasing magnetic field.

The top-right panel of \fref{fig:chrom_evolution} shows the evolution of the chiral charge density $\overline{n}_5(t)$, calculated using \eref{eq:n5_and_h}, and the magnetic helicity density $\overline{h}(t)$, calculated using \eref{eq:chrom_helicity}.  
These quantities appear to be constant at early time ($t \lesssim 10 t_\CPI{}(t_i))$, they abruptly change at $t \approx 10 t_\CPI{}(t_i)$ with the chiral density decreasing by a factor of $\approx 1/2$, and then they approach $\overline{n}_5 \to 0$ and $\tfrac{\alpha}{\pi} \overline{h} \to \overline{n}_5(t_i)$ at late time. In addition, $\overline{n}_5 \to 0$ as approximately $t^{-1/2}$ at late time.
The figure also shows the constant sum $\overline{n}_5 + 4 \tfrac{\alpha}{4\pi} \overline{h}$, which is consistent with the analytical discussion in \sref{sub:mono_spectrum}.  

The bottom-left panel of \fref{fig:chrom_evolution} shows the evolution of the various energy components of the system: the thermal energy $\overline{\rho}_T(t)$, the electromagnetic energy $\overline{\rho}_\mathrm{em}(t)$, the chiral energy $\overline{\rho}_5(t)$, and their sum the total energy $\overline{\rho}_\mathrm{tot}(t)$.  
These quantities are calculated using \eref{eq:rho_def}, except for the electromagnetic energy, which is calculated using $\overline{\rho}_\mathrm{em} = \tfrac{1}{2} \langle \Evec \cdot \Evec \rangle + \tfrac{1}{2} \langle \Bvec \cdot \Bvec \rangle$ and \eref{eq:exp_XY}.  
These energies appear to remain constant at early times, they change abruptly at $t \approx 10 t_\CPI{}(t_i)$, and they appear to approach constant asymptotic values at late times.  
The chiral energy decreases as the chiral asymmetry is consumed, the electromagnetic energy grows as the \CPI{} develops, and the thermal energy also grows as the excess chiral energy is expended.  
However, toward late times the electromagnetic energy again decreases toward zero.  
In the next subsection, we'll discuss how this is a consequence of Ohmic dissipation, which tends to deplete the field strength on small length scales.\footnote{
Since we are restricting the fluid to remain at rest, \ie{} $\uvec(\xvec,t) = 0$, our field simply depletes.  However, when fluid motion is taken into account, we anticipate that a turbulent cascade would occur, thereby sustaining the magnetic field.} 
Consequently, almost all of the initial chiral energy is transformed into thermal energy with the total energy remaining constant. 

The bottom-right panel of \fref{fig:chrom_evolution} shows the evolution of the plasma temperature $\overline{T}(t)$ relative to its initial value, $\dT(t) = \overline{T}(t) - T_i$. 
The temperature appears to be constant at early times, it rises abruptly at $t \approx 10 t_\CPI{}(t)$, and it appears to asymptote toward a constant $\dT \approx 10^{-5} T_i$ at late times.  
Since the fractional change remains small $\dT / T_i \ll 1$, the temperature change is proportional to the thermal energy change $\delta \rho_T \propto \dT$, which appears in the bottom-left panel. 
Because approximately all of the energy stored in the chiral asymmetry is eventually converted into thermal energy, the temperature change saturates to  
\bes{\label{eq:chromo_dT}
    \dT & = \frac{15 \mu_i^2}{4\pi^2 g_E T_i} 
    \com
}
which we have drawn in the bottom-right panel of \fref{fig:chrom_evolution} as the blue-dashed line.
We observe that the saturated temperature change agrees well with the monochromatic analytical expression in \eref{eq:dT_mono} in the limit $|\kvec_\star| \to 0$. 
Physically, the limit $|\kvec_\star| \to 0$ here is equivalent to the non-monochromatic system's behavior in the limit $t\to\infty$ as shorter wavenumbers correspond to longer timescales.

\begin{figure}[t]
    \centering
    \includegraphics[width=1.0\columnwidth]{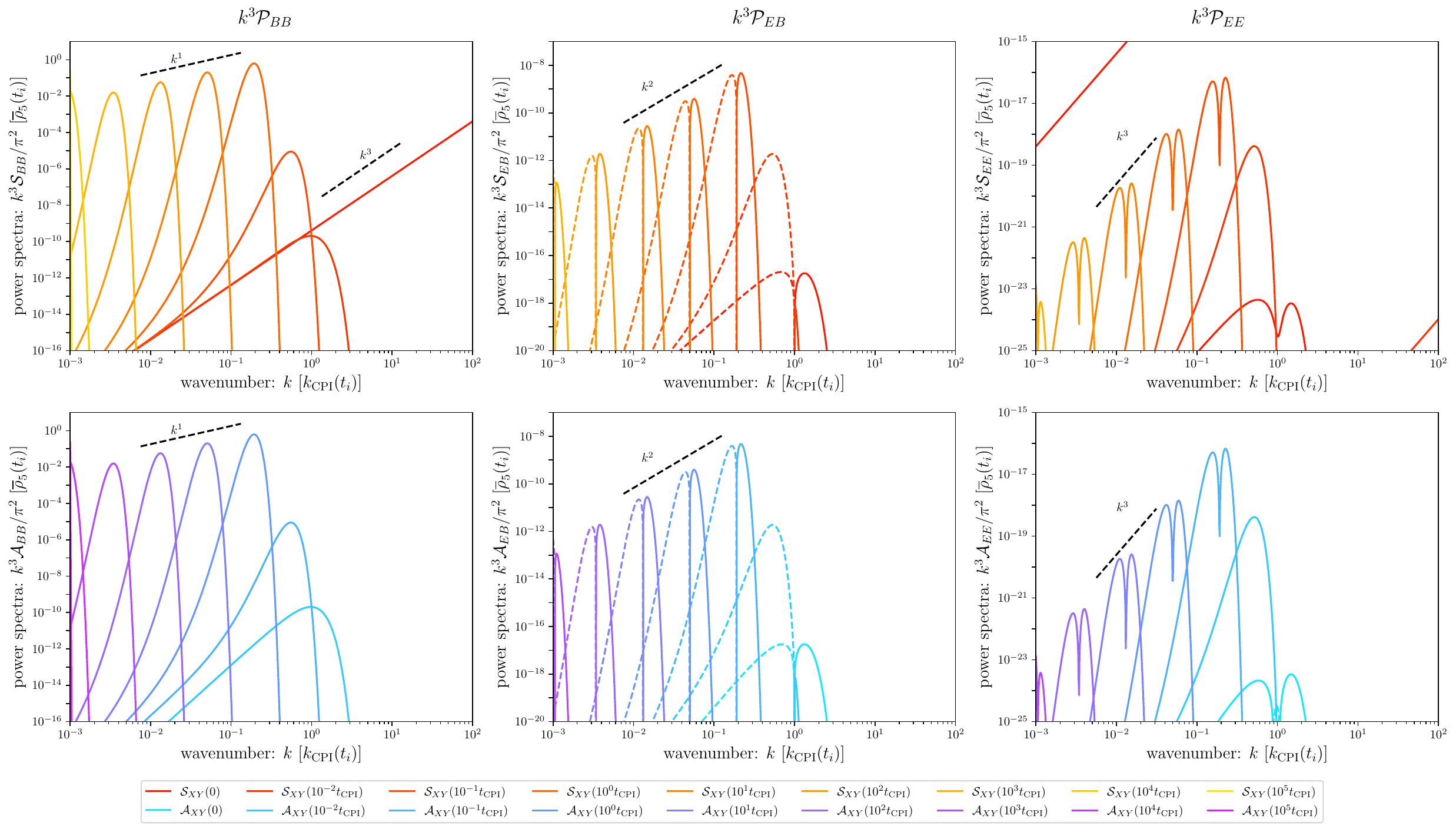} 
    \caption{
    \label{fig:chrom_spectra}
    Evolution of the various power spectra with time.  The top row shows symmetric power spectra $\mathcal{S}_{XY}(k,t)$, and the bottom row shows asymmetric power spectra $\mathcal{A}_{XY}(k,t)$.  The three columns correspond to $XY = EE$, $EB$, and $BB$, respectively.  Note that the vertical scale differs from panel to panel.  Each curve corresponds to a different time, shown on the legend.  If the function takes negative values, we draw it with a dashed line.
    In the top-right panel, the top straight line corresponds to the initial condition.
    }
\end{figure}

\subsubsection{Evolution of spectra}
\label{sub:spectra_evolution}

In \fref{fig:chrom_spectra}, we show the evolution of the various electromagnetic power spectra.  
In this section, we use $\mathcal{P}_{XY}(k,t)$ to refer to the symmetric power spectra $\mathcal{S}_{XY}(k,t)$ and the antisymmetric power spectra $\mathcal{A}_{XY}(k,t)$, collectively.  

The top-left and bottom-left panels of \fref{fig:chrom_spectra} show the magnetic field power spectra, $\mathcal{S}_{BB}(k,t)$ and $\mathcal{A}_{BB}(k,t)$ respectively.  
At the initial time $t = t_i = 0$, the initial conditions \pref{eq:chrom_initial} imply $k^3 \mathcal{S}_{BB} \propto k^3$ (since $k \ll T_i$ for the range of wavenumbers shown) and $\mathcal{A}_{BB} = 0$.  
At early times $t \lesssim t_\CPI{}(t_i)$ before the instability develops, the symmetric power spectrum $\mathcal{S}_{BB}(k,t)$ decreases with time in its high-$k$ modes, and it remains approximately constant in its low-$k$ modes.  
This behavior is a consequence of Ohmic dissipation, which can be understood using \eref{eq:eqns_of_motion_chromatic}.  
Since the electric conductivity is high, it is a good approximation to take $\frac{\partial}{\partial t} \mathcal{P}_{EB} \approx 0$ and  $\mathcal{P}_{EE} \approx 0$.  
Then, for modes $k \gg \overline{\mu}_5$, the equations of motion admit a solution 
\bes{\label{eq:SBB_approx_SBBi_exp}
    \mathcal{S}_{BB}(k,t) \approx \mathcal{S}_{BB,i}(k) \, \ee^{-2 [k/k_d(t)]^2} 
    \com
}
where $k_d(t) = \sqrt{\overline{\sigma}/t}$ sets the scale of Ohmic dissipation.  
Modes with $k \gtrsim k_d(t)/\sqrt{2}$ experience an exponential suppression, while modes with $k \lesssim k_d(t)/\sqrt{2}$ remain approximately unchanged. 
Meanwhile, the asymmetric power spectrum $\mathcal{A}_{BB}(k,t)$ grows from zero, and begins to track the symmetric power spectrum $\mathcal{S}_{BB}(k,t)$.  
This behavior can be understood from \eref{eq:eqns_of_motion_chromatic}; using the same approximations mentioned above \eref{eq:SBB_approx_SBBi_exp} leads to
\bes{ \label{eq:ABB_develop}
    \frac{\partial}{\partial t} \mathcal{A}_{BB}(k,t) 
    \approx 
    - 2 \frac{k^2}{\overline{\sigma}} \mathcal{A}_{BB}
    + 2\frac{k k_\CPI{}}{\overline{\sigma}} \mathcal{S}_{BB} 
    \per
}
Initially, $\mathcal{A}_{BB} \ll \mathcal{S}_{BB}$, and the symmetric power spectrum sources the asymmetric one.  
Later, the asymmetric power spectrum rises to $\mathcal{A}_{BB} \approx (k_\CPI{}/k) \, \mathcal{S}_{BB}$, and subsequently they evolve together. 
The \CPI{} begins to develop when $t \approx 10 t_\CPI{}(t_i)$, and the figure shows a sudden rise in both $\mathcal{S}_{BB}$ and $\mathcal{A}_{BB}$, particularly for the modes with $k \approx k_\CPI{}(t_i) / 2$.  
Modes with $k \ll k_\CPI{}(t_i)$ grow more slowly, and modes with $k \gtrsim k_d(t)/\sqrt{2}$ are suppressed by Ohmic dissipation.  
Since the \CPI{} depletes the chiral asymmetry, the instability wavenumber $k_\CPI{}(t)$ becomes smaller.  
Consequently the maxima of the power spectra shift toward smaller wavenumbers; we observe that peak location scales as $k = k_\mathrm{peak}(t) \propto t^{-1/2}$ and the peak height scales as $k^3 \mathcal{P}_{BB}(k,t)\big|_{k \approx k_\mathrm{peak}} \propto k_\mathrm{peak} \propto t^{-1/2}$.  
One can understand this behavior by noting that the helicity spectrum in \eref{eq:chrom_helicity} is proportional to $k^2 \mathcal{A}_{BB}(k,t)$.  
In order for helicity to be approximately constant at late times, as seen in \fref{fig:chrom_evolution}, its spectrum must satisfy $k^2 \mathcal{A}_{BB}(k,t) \bigr|_{k \approx k_\mathrm{peak}} \propto t^0$, which leads to the relation for $k^3 \mathcal{P}_{BB}$ given above. 
We compare this result with that of \rref{Sigl:2015xva}, which performed simulations of the chiral magnetic effect. 
The authors simulate the chiral magnetic effect for both a uniform spectral distribution and a Kolmogorov spectral distribution in a neutron star. 
They observed the same $k_\mathrm{peak}\propto t^{-1/2}$ behavior in their numerical simulations. In addition, they observe this behavior to be largely independent of initial conditions at late time.

The top-middle and bottom-middle panels of \fref{fig:chrom_spectra} show the cross field power spectra, $\mathcal{S}_{EB}(k,t)$ and $\mathcal{A}_{EB}(k,t)$.  
Note that the cross spectra can be negative (indicated by dashed curves). 
As only positive helicity modes are amplified in this system, $\mathcal{S}_{EB}(k,t) = \mathcal{A}_{EB}(k,t)$.
We observe that $\mathcal{P}_{EB}(k,t)$ may change sign as $k$ is varied.  
This behavior can be understood from the equations of motion \pref{eq:eqns_of_motion_chromatic} with the approximations $\mathcal{A}_{XY} \approx \mathcal{S}_{XY}$ and $\mathcal{P}_{EB} \approx \mathcal{P}_{EE} \approx 0$, which imply 
\bes{
    \frac{\partial}{\partial t} \mathcal{P}_{EB} 
    \approx (k - k_\CPI{}) \, \mathcal{P}_{BB} - \overline{\sigma} \, \mathcal{P}_{EB}
    \per
}
This relation reveals how the magnetic power spectra $\mathcal{P}_{BB}$ source the cross power spectra $\mathcal{P}_{EB}$.  
For $k > k_\CPI{}(t)$, the source is positive leading to $\mathcal{P}_{EB} > 0$, and for $k < k_\CPI{}(t)$, the source is negative leading to $\mathcal{P}_{EB} < 0$.  
If the source is not sufficiently large in magnitude, \ie{} $|k-k_\CPI{}| | \mathcal{P}_{BB}(k,t)| < \overline{\sigma} |\mathcal{P}_{EB}(k,t)|$, then the power is exponentially depleted $\propto \ee^{-\overline{\sigma} t}$ due to Ohmic dissipation.  
After the \CPI{} develops, the maxima of the power spectra shift toward smaller wavenumbers; we observe that the peak location scales as $k = k_\mathrm{peak} \propto t^{-1/2}$ and the peak height scales as $k^3 \mathcal{P}_{EB}(k,t) \big|_{k \approx k_\mathrm{peak}} \propto k_\mathrm{peak}^2 \propto t^{-1}$.  
This behavior can be understood from the relations noted earlier: $\tfrac{\partial}{\partial t} \mathcal{P}_{EB} \sim k \, \mathcal{P}_{BB}$ and $k^3 \mathcal{P}_{BB} \big|_{k \approx k_\mathrm{peak}} \propto k_\mathrm{peak} \propto t^{-1/2}$.  
The top-right and bottom-right panels of \fref{fig:chrom_spectra} show the electric field power spectra, $\mathcal{S}_{EE}(k,t)$ and $\mathcal{A}_{EE}(k,t)$.  
We observe that both $\mathcal{S}_{EE}(k,t)$ and $\mathcal{A}_{EE}(k,t)$ pass through zero as $k$ is varied.  
This behavior can be understood from the equation of motion \pref{eq:AEE}.
The first term reveals that $\mathcal{S}_{EB}$ sources $\mathcal{A}_{EE}$. 
At early times, the sign of $\mathcal{A}_{EE}$ tracks the sign of $\mathcal{S}_{EB}$.  
At later times, the zero-crossings of $\mathcal{S}_{EB}$ control the zeros of $\mathcal{A}_{EE}$, since it is the electric field that vanishes at $k \approx k_\CPI{}$.
In addition, the last term in \eref{eq:SEE} shows how the electric field is initially depleted.  
This occurs because initially $\mathcal{P}_{EB}=0$ and therefore the last term in \eref{eq:SEE} causes $\mathcal{S}_{EE}$ to exponentially decrease.
After the \CPI{} develops, the maxima of the power spectra shift toward smaller wavenumbers; we observe that the peak location scales as $k = k_\mathrm{peak} \propto t^{-1/2}$ and the peak height scales as $k^3 \mathcal{P}_{EE}(k,t) \big|_{k \approx k_\mathrm{peak}} \propto k_\mathrm{peak}^3 \propto t^{-3/2}$.  
This behavior can be understood from the relations noted earlier: $\tfrac{\partial}{\partial t} \mathcal{P}_{EE} \sim k \, \mathcal{P}_{EB}$ and $k^3 \mathcal{P}_{EB} \big|_{k \approx k_\mathrm{peak}} \propto k_\mathrm{peak}^2 \propto t^{-1}$.

\begin{figure}[t]
    \centering
    \includegraphics[width=0.75\columnwidth]{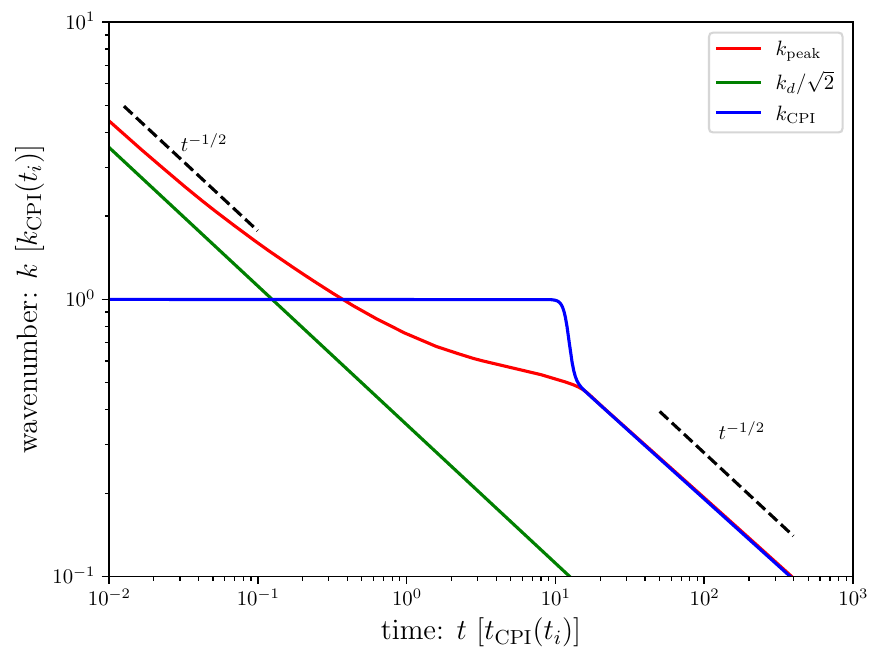}
    \caption{
    \label{fig:wavenumbers}
    Evolution of relevant length scales.  We plot the wavenumber $k = k_\mathrm{peak}(t)$ at which $k^3 \mathcal{S}_{BB}(k,t)+k^3 \mathcal{S}_{EE}(k,t)$ is maximized at time $t$, the wavenumber $k = k_d(t)/\sqrt{2}$ above which Ohmic dissipation is relevant, and the wavenumber $k = k_\CPI{}(t)$ below which the \CPI{} is active. 
    }
\end{figure}

Finally, in \fref{fig:wavenumbers} we summarize the various relevant length scales. 
We show the wavenumber $k_d(t)/\sqrt{2} = \sqrt{\overline{\sigma}/2t}$, above which Ohmic dissipation leads to an exponential depletion of the field amplitude.  
Since $\overline{\sigma}(t) = 100 \overline{T}(t)$ remains approximately constant (because the heating is negligible for our chosen parameters) this wavenumber scales as $k_d(t) \propto t^{-1/2}$.  
We show the wavenumber $k_\CPI{}(t) = \tfrac{2\alpha}{\pi} \overline{\mu}_5(t)$, below which the \CPI{} is active.  
Before the instability develops, the chiral asymmetry is approximately constant and equal to its initial value $\overline{\mu}_5(t) \approx \mu_i$, which implies $k_\CPI{}(t) \approx \tfrac{2\alpha}{\pi} \mu_i$ is approximately constant.  
After the instability develops at time $t \approx 10 t_\CPI{}(t_i)$, the chiral asymmetry suddenly drops by a factor of $\approx 1/2$ and then continues to decrease toward zero, which gives the scaling $k_\CPI{}(t) \propto \overline{\mu}_5(t) \propto t^{-1/2}$.  
We show the wavenumber $k_\mathrm{peak}(t)$, which is defined as the maximum of $k^3 \mathcal{S}_{BB}(k,t)+k^3 \mathcal{S}_{EE}(k,t)$ (though the $k^3 \mathcal{S}_{BB}(k,t)$ term dominates). 
Physically, $k_\mathrm{peak}(t)$ is the wavenumber where most of the magnetic energy is stored.  
Before the instability develops, this scale evolves like $k_\mathrm{peak} \propto t^{-1/2}$, which tracks $k_d(t) \propto t^{-1/2}$, since Ohmic dissipation depletes magnetic field modes with $k > k_d/\sqrt{2}$.  
After the instability develops, this scale evolves to track the instability scale, \ie{}, $k_\mathrm{peak}(t) \approx k_\CPI{}(t) \propto t^{-1/2}$, because the magnetic field dominates the electromagnetic energy spectrum.
This agrees with the discussion of the simulations in \rref{Sigl:2015xva} as discussed below \eref{eq:ABB_develop}.

\subsubsection{Large chemical potential}
\label{sub:large_mu}

We recall that earlier in this section, \eref{eq:chromo_dT} shows that for small $\mu_i/T_i$, the heating of the plasma goes like $\dT \propto \mu_i^2/T_i$. One may then ask how much would the plasma be heated if $\mu_i\sim T_i$ or $\mu_i \gg T_i$. As one may guess from \eref{eq:rho_5}, in such a regime, the energy stored in the chiral asymmetry would be:
\ba{
    \overline{\rho}_5(t) = \tfrac{1}{2} \overline{\mu}_5^2 \overline{T}^2 + \tfrac{1}{4\pi^2} \overline{\mu}_5^4. 
}
As with small $\mu_i/T_i$, for large $\mu_i/T_i$, the chiral anomaly depletes $\overline{\mu}_5$ in order to produce the helical magnetic field. 
However, since the chiral plasma instability requires a seed magnetic field in order to produce a magnetic field and for a thermal distribution, modes with wavenumber $k>T_i$ are initially exponentially suppressed, for $\mu_i \gg T_i$ one may question if the instability is able to develop as $k_\CPI{}(t_i) = 2 \alpha \mu_i/\pi \gg T_i$. 
To resolve this, we argue that the wavenumbers that satisfy $k\ll \overline{T}$ will grow by depleting $\overline{\mu}_5$ similar to the $\mu_i \ll T_i$ case.
As a result, even if initially $\mu_i\gg T_i$, the system will eventually reach a regime where $\overline{\mu}_5\ll \overline{T}$ and the behavior discussed early in this section would be regained. 
One may also question if the constitutive relation holds for $\mu_i \gg T_i$ as the length scale for the chiral plasma instability $l_\CPI{}(t_i)\equiv2\pi/k_\CPI{}(t_i)$ is much less than $T_i^{-1}$, which, for $\mu_i \ll T_i$, describes the distance between the fermions in the plasma ($l_n$).
We address this concern by noting that the general expression for the chiral number density $\overline n_5(t)$ for an arbitrary $\overline \mu_5$ is
\begin{align}
    \overline{n}_5 = \frac{1}{3} \overline{\mu}_5 \overline {T}^2 + \frac{1}{3\pi^2}\overline{\mu}_5^3,
\end{align}
which for $\overline \mu_5 \gg \overline{T}$ simplifies to $\overline{n}_5 \approx \overline{\mu}_5^3/(3\pi^2)$. 
From this, we can see the length scale for the distance between the fermions in the plasma goes like $l_n \sim \overline \mu_5^{-1}$ for $\overline \mu_5 \gg \overline{T}$. As $k_\CPI{} \sim \alpha \overline{\mu}_5$ and $\alpha \ll 1$, for $\overline \mu_5 \gg \overline{T}$, $l_n \ll l_\CPI{}$ and thus the constitutive relation holds.

\begin{figure}[t]
    \centering
    \includegraphics[width=0.75\columnwidth]{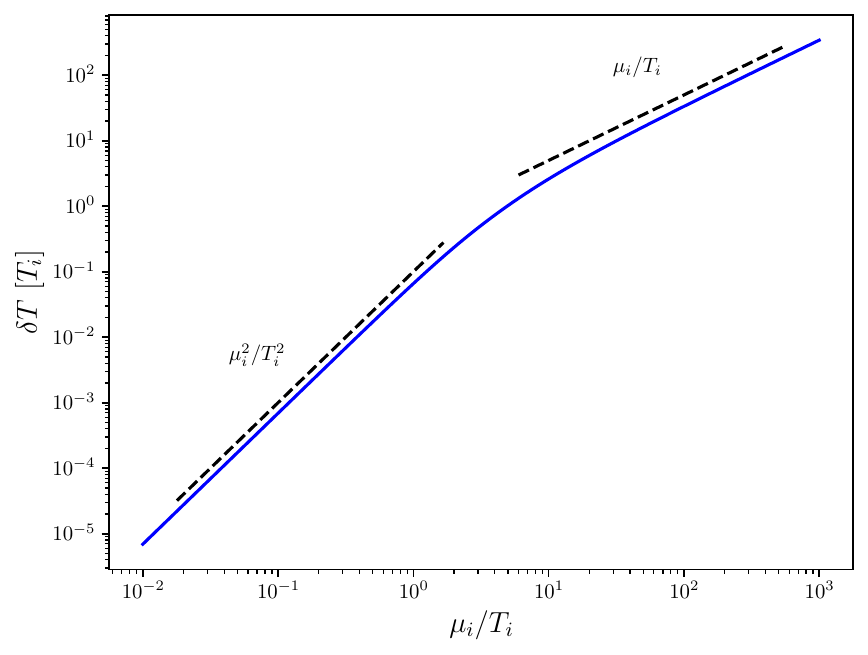}
    \caption[Temperature Evolution for Large $\mu_i$ ]{\label{fig:large_mu}
    Predicted heating for both small and large chiral asymmetry.  We show the temperature increase $\dT$, given by \eref{eq:gen_delta_T}, across a range of values for the initial chiral chemical potential $\mu_i = \mu_5(t_i)$, corresponding to small chiral asymmetry ($\mu_i/T_i \ll 1$) and large chiral asymmetry ($\mu_i/T_i \gg 1$).  
    }
\end{figure}

To determine how much the plasma heats, we observe that at sufficiently late time, Ohm's law depletes $k$ modes until the energy in the fields is negligible. As a result, all of the energy of the system eventually is transferred to the temperature of the plasma in the form of heat. Therefore, as the initial energy in the fields is also negligible, using conservation of energy, one can observe that:
\ba{
    \overline{\rho}_5(t_i) +\overline{\rho}_T(t_i) \approx\overline{\rho}_T(t_f),
}
where $t_f$ is defined to be at sufficiently late time that the system has stopped evolving significantly.  
As a result, one can see that for an arbitrary $\mu_i/T_i$, the change in the temperature of the plasma due to the chiral magnetic effect can be calculated to be: 
\bes{\label{eq:gen_delta_T}
    \dT \approx \biggl( \frac{15^{1/4}}{2^{1/4} g_E^{1/4} \pi } \mu_i \biggr) 
    \biggl[ \Big( 1 + 2 \pi^2 \frac{T_i^2}{\mu_i^2} + \frac{2}{15} g_E \pi^4 \frac{T_i^4}{\mu_i^4} \Big)^{1/4} - \frac{2^{1/4} g_E^{1/4} \pi}{15^{1/4}} \frac{T_i}{\mu_i} \biggr].
}
In \fref{fig:large_mu}, we have plotted the change in temperature $\dT$ for a range of $\mu_i/T_i$. This has been done by plotting \eref{eq:gen_delta_T}. From \fref{fig:large_mu}, we can see that for $\mu_i\ll T_i$, $\dT \propto \mu_i^2$ as before. However, for  $\mu_i\gg T_i$ we can see that $\dT \propto \mu_i$. To explain this,
we can see that for $\mu_i\gg T_i$, in terms in the square brackets in \eref{eq:gen_delta_T} become approximately 1, and thus the change in temperature is $\dT \approx \bigl( \tfrac{15}{2 g_E \pi^4} \bigr)^{1/4} \mu_i $.
From this we are able to see that $\dT \propto \mu_i$ is the expected behavior for large $\mu_i/T_i$.

\section{Conclusion}
\label{sec:conc}

We have investigated the energetics of a relativistic chiral plasma, which is often studied in the context of early universe cosmology, compact stars, and heavy ion collisions.  
It is well known that a nonzero chiral asymmetry is expected to open a tachyonic instability (\CPI{}) towards the growth of a helical magnetic field and the depletion of the chiral charge.  
The work reported in this article was motivated by a desire to understand how energy is transferred from the fluid to the electromagnetic field. 

Since there is an energy associated to the chiral asymmetry itself, $\rho_5 \sim \mu_5^2 T^2$, one might expect that the growth in the electromagnetic energy $\rho_\mathrm{em} \sim |\Bvec|^2$ is exactly compensated by the depletion of the chiral energy, \ie{} $\Delta \rho_\mathrm{em} = - \Delta \rho_5$.  
Somewhat surprisingly, we find that this is not the case.  
Instead the transfer of energy from the fluid to the field is accompanied by a heating. 

\begin{figure}
    \centering
    \includegraphics[width=0.75\linewidth]{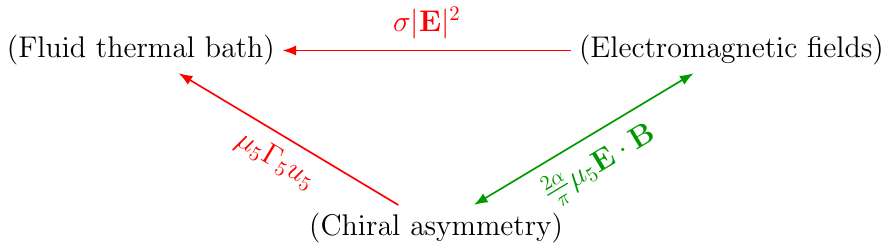}
    \caption{Energy flow among the various energy reservoirs of the \cMHD{} system. Red arrows indicate the irreversible, thus entropy producing, processes due to Ohmic dissipation ($\sigma|\Evec|^2$) and chirality-flipping scatterings ($\mu_5\Gamma_5 n_5$), whereas the green double arrow indicates the reversible, thus not entropy-producing, interaction due to the chiral anomaly ($\tfrac{2\alpha}{\pi}\mu_5\Evec\cdot\Bvec$).}
    \label{fig:energy_flow}
\end{figure}

In order to study energy transfer, we adopt the theory of chiral magnetohydrodynamics.  
However, the ``standard'' equations of \cMHD{} proved to be inadequate for our purposes.  
We show that these equations are not consistent with energy conservation at order $\mu_5/T$.  
Although the \cMHD{} equations can be derived by imposing covariant energy-momentum conservation, in the course of this derivation, several typically-small $O(\mu_5/T$) terms are neglected, thereby leading to the ``standard'' \cMHD{} equations that are not consistent with exact energy conservation. 
By restoring the neglected terms, we arrive at a system of extended \cMHD{} equations of motion \pref{eq:cMHD_corrected} that incorporate exact energy conservation and temperature change. 
Although it is only an intermediate step in our effort to understand energy transfer, we view these equations as one of the main results of our work. 
The flow of energy among the various components of the system (fluid thermal bath, electromagnetic fields, and the chiral asymmetry) is visualized in the summary figure \ref{fig:energy_flow}. 
The irreversible thermodynamic processes due to Ohmic dissipation and chirality-flipping scatterings (indicated with red arrows in the figure) produce a positive entropy injection into the thermal background plasma.
We have neglected $\Gamma_5$ in the analysis of the extended \cMHD{} equations, correspondingly, the heating of the plasma was only indirectly sourced by the asymmetry through the resistive damping of the electromagnetic fields.

Using our extended \cMHD{} equations, we employ a combination of analytical and numerical methods to calculate how the fluid and field variables evolve during the \CPI{}.  
For the sake of concreteness and to make the calculations analytically tractable, we consider two simple systems in which the plasma is homogeneous and at rest.  
The \CPI{} would be more accurately modeled by allowing for field and fluid inhomogeneities with nonzero bulk velocity, but we do not expect that doing so would change our conclusions regarding energy transfer and heating.  
That is, we expect the parametric relationship $\dT \sim \mu_5^2 / T$ to be unchanged, but the numerical coefficient may be impacted.  
A natural extension of our work would be a study that numerically solves our extended \cMHD{} equations to calculate the amount of heating while accounting for inhomogeneities and nonzero bulk velocity. 

In the two simple systems that we study, we find that the chiral asymmetry carries more energy than is used to create the growing helical magnetic field, and that the additional energy leads to an increase in the temperature of the plasma.  
For a small initial chiral asymmetry, \ie{} $\mu_i = \mu_5(t_i) \ll T_i = T(t_i)$, the temperature increase is parametrically $\dT \sim \mu_i^2 / T_i$ as shown in \eref{eq:chromo_dT}.  
To the best of our knowledge, \CPI{}-induced heating was overlooked in previous studies of chiral plasmas.  
Presumably, this is because the heating is typically insignificant, since $\dT / T_i \ll 1$ for $\mu_i / T_i \ll 1$.  
Further work is needed to systematically extend our analysis into the regime $\mu_5 / T > 1$, for example by accounting for $\mu_5$-dependence in the transport coefficients.  
However, we expect that systems with a large chiral asymmetry would experience an appreciable heating.  
Degenerate systems with large $\mu/T$ have been considered in the context of early universe cosmology~\cite{Carena:2021bqm,Batell:2024hzo} and astrophysics~\cite{Tremaine:1979we,Bai:2018dxf,Hong:2020est,Chavanis:2021jds}.  

If an appreciable heating were to occur in the early universe, it could leave an imprint on various cosmological relics.  
Heating during the epoch of nucleosynthesis would impact the abundances of light elements, possibly disrupting the excellent agreement between the theory of Big Bang Nucleosynthesis and the observed abundances of light elements \cite{Kawasaki:2017bqm}.
Heating during the epoch prior to recombination would impact the energy distribution of the cosmic microwave background radiation, possibly inducing a spectral distortion that could be identified with future precision measurements of the nearly black body spectrum~\cite{Kogut:2019vqh}. 
Heating between nucleosynthesis and recombination is also strongly constrained~\cite{Sobotka:2022vrr,Sobotka:2023bzr}.  
However, in our scenario the \CPI{}-induced heating would take place at $T \gtrsim 80 \; \mathrm{TeV}$, long before nucleosynthesis and recombination. 
Nevertheless, an appreciable heating would modify the cosmological expansion rate~\cite{Batell:2024dsi}, which leaves a distinctive imprint on the spectrum of  primordial gravitational waves if they are produced before the heating occurs~\cite{Cui:2017ufi,Cui:2018rwi,Gouttenoire:2021jhk}.

\acknowledgments
This material is based upon work supported (in part: B.A. and A.J.L.) by the National Aeronautic and Space Administration (NASA) Astrophysics Theory (ATP) Program award 80NSSC22K0825.  
This research was supported by the Deutsche Forschungsgemeinschaft (DFG, German Research Foundation) under Germany’s Excellence Strategy– EXC 2121 “Quantum Universe”– 390833306.
This research was funded (in part: K.S.) by the Excellence Programme of the Hungarian Ministry of Culture and Innovation grant number TKP2021-NKTA-64 and by the Hungarian Scientific Research Fund grant number PD-146527.

\appendix

\section{Relativistic MHD equations}
\label{app:RelativisticMHD}
In \sref{sec:eqns_of_chiMHD}, we presented the standard \cMHD{} equations in the non-relativistic limit where higher powers of the bulk velocity $\uvec$ were neglected.
As pointed out recently in \cite{RoperPol:2025lgc}, these equations result from an inconsistent limit where spatial and temporal derivatives of the Lorentz $\gamma$ factor are dropped, even though they give leading order terms.
In this appendix, we present the full equations and their appropriate non-relativistic limits.
It should be emphasized that the modifications resulting from the proper relativistic limit to the standard \cMHD{} equations have no bearing on the arguments for energy-momentum conservation and thermodynamics, as were discussed in \sref{sec:chiral_MHD}.

To obtain the energy-momentum continuity equation, we start from \eref{eq:continuity_eq} with the definitions of the stress-energy tensors in \erefs{eq:T_munu_fl}{eq:T_munu_em}, and using the four-velocity $u^\mu = \gamma(|\uvec|)(1,\uvec)$.
Performing the temporal and spatial derivatives, one finds the temporal component of the energy-momentum continuity equation, $\partial_\mu T^{\mu 0}=0$,
\begin{align}
    \label{eq:continuity_full_temporal}
    {\tfrac{\partial}{\partial t}\rho_{\rm fl}} = - \tfrac{4\gamma^2}{4\gamma^2-1}\dvec\cdot(\rho_{\rm fl}\uvec) + \tfrac{3}{4\gamma^2-1}\Evec\cdot\Jvec - \tfrac{8\gamma^4}{4\gamma^2-1}\rho_{\rm fl} \uvec\cdot\big[\tfrac{\partial}{\partial t}\uvec + (\uvec\cdot\dvec)\uvec\big]\com
\end{align}
and the spatial part of the energy-momentum continuity equation, $\partial_\mu T^{\mu i}=0$,
\begin{equation}
\begin{aligned}
    \label{eq:continuity_full_spatial}
    &\rho_{\rm fl} \big[\tfrac{\partial}{\partial t}\uvec + (\uvec\cdot\dvec)\uvec\big] = \\ &\qquad \tfrac{3}{4\gamma^2}\Jvec\times\Bvec - \tfrac{1}{4\gamma^2}\dvec\rho_{\rm fl} - \uvec\Big[{\tfrac{\partial}{\partial t}\rho_{\rm fl}} + \dvec\cdot(\rho_{\rm fl}\uvec) +   2\gamma^2\rho_{\rm fl} \uvec\cdot\big[\tfrac{\partial}{\partial t}\uvec + (\uvec\cdot\dvec)\uvec\big]\Big] \per
\end{aligned}
\end{equation}
These equations are identical to those presented originally in eqs. (9) and (10) in ref.~\cite{Brandenburg:1996fc}.
The last step to obtain the MHD equations is diagonalizing \erefs{eq:continuity_full_temporal}{eq:continuity_full_spatial} for $\tfrac{\partial}{\partial t}\rho_{\rm fl}$ and $\rho_{\rm fl}\big[\tfrac{\partial}{\partial t}\uvec + (\uvec\cdot\dvec)\uvec\big]$.
This is most easily done by splitting the vectorial terms into components that are parallel or perpendicular to $\uvec$.
The energy continuity equation becomes:
\begin{equation}
\label{eq:drhofldt_relat}
    \tfrac{\partial}{\partial t} \rho_{\rm fl} = \tfrac{3(2\gamma^2-1)}{2\gamma^2+1}\Evec\cdot\Jvec - \tfrac{6\gamma^2}{2\gamma^2+1}\uvec\cdot(\Jvec\times\Bvec) - \tfrac{4\gamma^2}{2\gamma^2+1}\rho_{\rm fl}\dvec\cdot\uvec - \tfrac{2\gamma^2}{2\gamma^2+1}\uvec\cdot\dvec\rho_{\rm fl} \com
\end{equation}
whereas the momentum continuity equation is:
\begin{equation}
\label{eq:Du_relat}
\begin{aligned}
    \rho_{\rm fl} \big[\tfrac{\partial}{\partial t}\uvec + (\uvec\cdot\dvec)\uvec\big] = &-\tfrac{3}{2\gamma^2+1}\uvec(\Evec\cdot\Jvec) + \tfrac{1}{2\gamma^2+1}\rho_{\rm fl}\uvec(\dvec\cdot\uvec) + \tfrac{3}{4\gamma^2}\Jvec\times\Bvec \\ & + \tfrac{3}{2(2\gamma^2+1)}\uvec[\uvec\cdot(\Jvec\times\Bvec)] - \tfrac{1}{4\gamma^2}\dvec\rho_{\rm fl} + \tfrac{1}{2(2\gamma^2+1)}\uvec[\uvec\cdot\dvec\rho_{\rm fl}] \per
\end{aligned}
\end{equation}
These equations are identical to those found in eq.~(6.31) of ref.~\cite{RoperPol:2025lgc}.

The non-relativistic limit of the \cMHD{} equations is obtained from \erefs{eq:drhofldt_relat}{eq:Du_relat} by dropping terms that are sub-leading in bulk speed $|\uvec|$ of the plasma.
For the energy continuity equation, we find
\begin{equation}
    \label{eq:drhofldt_relat_LO}
    \tfrac{\partial}{\partial t}\rho_{\rm fl} = \Evec\cdot\Jvec - 2\uvec\cdot(\Jvec\times\Bvec) - \frac{4}{3} \rho_{\rm fl} \dvec\cdot\uvec - \frac{2}{3}\uvec\cdot\dvec\rho_{\rm fl} + \mathcal{O}(|\uvec|^2)\com
\end{equation}
which differs from the standard \cMHD{} form in \eref{eq:drhofldt} and
the energy conserving form we derived in \eref{eq:MHD_drhofl_dt}, although, it matches the latter when $\uvec=0$.
Similarly, for the momentum continuity equation we find
\begin{equation}
\label{eq:Du_relat_LO}
\begin{aligned}
    \rho_{\rm fl} \big[\tfrac{\partial}{\partial t}\uvec + (\uvec\cdot\dvec)\uvec\big] = -\uvec(\Evec\cdot\Jvec) + \frac{1}{3}\rho_{\rm fl}\uvec(\dvec\cdot\uvec) + \frac{3}{4}(1 - |\uvec|^2) \Jvec\times\Bvec \cr + \frac{1}{2} \uvec\big[\uvec\cdot(\Jvec \times\Bvec) \big] - \frac{1}{4}(1 - |\uvec|^2) \dvec\rho_{\rm fl} + 
    \frac{1}{6}\uvec(\uvec\cdot\dvec\rho_{\rm fl}) + \mathcal{O}(|\uvec|^3)\per 
\end{aligned}
\end{equation}
This again is different from the previously presented equations in \erefs{eq:dudt}{eq:MHD_Du_dt}.

In summary, for the proper non-relativistic set of \cMHD{} equations that are consistent with energy-momentum conservation also, one must replace \erefs{eq:MHD_drhofl_dt}{eq:MHD_Du_dt} with those given in \erefs{eq:drhofldt_relat_LO}{eq:Du_relat_LO}, respectively.

\section{Late-time scaling of the chiral chemical potential}
\label{app:chemical_potential_scaling}

This appendix supplements the discussion of the non-monochromatic spectrum evolution in \sref{sub:chrom_spectrum}.  
We provide a careful derivation of the evolution of the chiral chemical potential $\overline\mu_5(t)$ after the initial instability has developed.

\subsection{Leading order behavior}
\label{app:leading_chemical_potential_scaling}

As discussed in \ref{sub:chrom_initial}, the leading order behavior of $\overline{\mu}_5(t)$ is $\overline{\mu}_5(t)\propto t^{-1/2}$. In this section we will derive this behavior. 
As shown in~\fref{fig:chrom_spectra}, the $\mathcal{P}_{BB}$ spectra's peaks are at $k \approx k_\CPI{}$ and at leading order, move to lower wavenumbers like $k_\CPI{}^1$. Therefore, we will impose the late time ansatz approximating the behavior near the peaks as:
\bes{ \label{eq:muA_behavior_start}
    \frac{k^3}{\pi^2}\mathcal{P}_{BB}(k,t)\big|_{k\approx k_\CPI{}(t)} \approx a(t) k_\CPI{}(t) - b(t) \frac{k^2}{k_\CPI{}(t)}[k-k_\CPI{}(t)],
}
where $a(t)$ and $b(t)$ are factors that we can calculate. Next, as the $k^3 \mathcal{\mathcal{P}}_{BB}(k,t)$ spectra are sharply peaked, we will approximate the behavior as being dominated by the region around $k_\CPI{}(t)$ such that one can  consider only the region of the spectra where $|k-k_\CPI{}|\ll  k_\CPI{}$.  
The magnetic helicity density $\overline{h}(t)$ is calculated by performing the integral in \eref{eq:chrom_helicity}. 
At late time, the integral is dominated by $k \approx k_\CPI{}(t)$ where our ansatz is a good approximation to the power spectrum.  
Using the ansatz and integrating across the range $k_\CPI{}/2 < k < 3 k_\CPI{}/2$ gives 
\bes{
    \overline{h} \approx  a(t)  - b(t)\frac{1}{12}k_\CPI{}.
}
Using the conservation of helicity and approximating the change in temperature as constant, one can then derive the relation:
\bes{
    \frac{1}{3}\mu_i T_i^2 &= \frac{1}{3} \overline{\mu}_5(t) T_i^2 + 4 \tfrac{\alpha}{4\pi}\Big(a   - b\frac{1}{12}k_\CPI{}(t)\Big)\\
    \Rightarrow a&=\frac{\pi\mu_i T_i^2}{3\alpha},\qquad b=\frac{2\pi^2 T_i^2}{\alpha^2}.
}
Using the equations of motion \pref{eq:eqns_of_motion_chromatic}, in the region $|k-k_\CPI{}|\ll  k_\CPI{}$, if we let $\mathcal{P}_{EE}\approx 0$ and $\frac{\dd}{\dd t} \mathcal{P}_{EB}\approx 0$, we can formulate the approximate relation:
\begin{align}
    \frac{k^3}{\pi^2}\mathcal{P}_{EB}(k,t)\big|_{k\approx k_\CPI{}} &\approx \frac{k-k_\CPI{}}{\overline{\sigma}}\mathcal{P}_{BB}(k,t) \big|_{k\approx k_\CPI{}}\\
    &\approx \tfrac{\pi\mu_i T_i^2}{3\alpha\overline{\sigma}} (k-k_\CPI{}) k_\CPI{}(t) - \tfrac{2\pi^2 T_i^2}{\alpha^2\overline{\sigma}}k(k-k_\CPI{}(t))^2 - \tfrac{2\pi^2 T_i^2}{\alpha^2\overline{\sigma}}\frac{k}{k_\CPI{}}(k-k_\CPI{}(t))^3.
\end{align}
As the terms odd in $k-k_\CPI{}$ cancel in the integral when calculating $\langle \Evec \cdot \Bvec\rangle$,
\bes{
    \langle \Evec \cdot \Bvec\rangle\approx -\frac{\pi^2 T_i^2 }{6\alpha^2\overline{\sigma}}k_\CPI{}^3.
}
Returning to~\eref{eq:eqns_of_motion_chromatic}, if we approximate $\overline{\mu}_5^4/\overline{T}^4 \approx  0$ and $\langle \Evec\cdot \Evec \rangle\approx 0$, the equation of motion for $\overline{\mu}_5$ becomes:
\bes{
    \tfrac{\dd}{\dd t} \overline{\mu}_5 
     \approx \frac{6 \alpha}{\pi} \frac{1}{\overline{T}^2} \langle \Evec \cdot \Bvec \rangle 
     = -\frac{\pi }{\alpha\overline{\sigma}}k_\CPI{}(t)^3 
     \propto - \overline{\mu}_5^3 
}
which has the solution, for large $t$, $\overline{\mu}_5\propto t^{-1/2}$ as found in the numerical solutions. This result is in agreement with the simulations and discussion in~\cite{Sigl:2015xva}.

\subsection{Correction to leading order behavior}

As presented at the end of section \ref{app:leading_chemical_potential_scaling}, the expected scaling of the chemical potential is $\overline\mu_5(t)\propto t^{-1/2}$.
However, while this provides a very good approximation, the numerical results favor a scaling that slightly differs from this, \ie{} $\overline\mu_5(t)\propto t^{-1/2}f(t)$ 
with $f(t)$ approximately constant at leading order in $t$. We shall derive an approximation of this correction here.

The full set of equations given in \eref{eq:eqns_of_motion_chromatic} is impossible to solve analytically, however, we can make a number of simplifying assumptions that allow for an analytical treatment.
As in the monochromatic case, whose analytical solution was discussed in \ref{sub:mono_initial}, we assume that 
(i) the temperature change provides only a sub-leading contribution to the evolution of the chemical potential such that $\overline{T}(t)\approx T_i$, 
(ii) the electric field is frozen (\ie{}, the ideal MHD limit with $\dot\Evec=0$), and
(iii) the evolution of the system is maximally helical.
From the full set of evolution equations for the spectrum in \sref{sub:chrom_spectrum}, using the above assumptions, we arrive at a set of two differential equations for the symmetric part of the magnetic spectrum and the chemical potential (written here for $k_\CPI{}(t)=2\alpha\overline\mu_5(t)/\pi$):
\begin{align}
    \label{eq:dSdt_app}
    \frac{{\rm d}}{{\rm d} t}\mathcal{S}_{BB}(|\kvec|,t) &= -\frac{2k^2}{\overline\sigma}\Big(1-\frac{k_\CPI{}(t)}{k}\Big)\mathcal{S}_{BB}(|\kvec|,t)\com \\
    \label{eq:dkAdt_app}
    \frac{{\rm d} k_\CPI{}(t)}{{\rm d} t} &= \frac{12\alpha^2}{\pi^2 T^2_i \overline \sigma}\int_0^\infty{\rm d} k\,k^3\mathcal{S}_{BB}(|\kvec|,t)\Big(1-\frac{k_\CPI{}(t)}{k}\Big)\per
\end{align}

As highlighted in \ref{sec:analytic_sol}, the differential equation in \eref{eq:dSdt_app} is formally solved as
\begin{equation}
    \label{eq:S_solution_analytic}
    \mathcal{S}_{BB}(|\kvec|,t) = S_i\exp\bigg[-\frac{2k\big(k-\overline k_\CPI{}(t)\big)}{\overline\sigma}t\bigg]\com
\end{equation}
where we approximate the initial spectrum as white noise 
with value $S_i$
and introduced the time-averaged chiral wavenumber
\begin{equation}
    \label{eq:kAbar_definition_app}
    \overline k_\CPI{}(t) = \frac{1}{t}\int_{0}^t{\rm d} t'\,k_\CPI{}(t')\per
\end{equation}
Next, notice that the right-hand side of \eref{eq:dkAdt_app} contains the right-hand side of \eref{eq:dSdt_app}; thus, by substituting the former into the latter,
one has the combined differential equation:
\begin{equation}
    \label{eq:kA_conservation_equation_app}
    \frac{{\rm d} k_\CPI{}(t)}{{\rm d} t} + \frac{6\alpha^2}{\pi^2 T^2_i}\frac{{\rm d}}{{\rm d} t}\int_0^\infty{\rm d} k\, k \mathcal{S}_{BB}(|\kvec|,t)=0\per
\end{equation}

The wavenumber integral is analytic for $\mathcal{S}_{BB}$ given in \eref{eq:S_solution_analytic}.
As the expression is a total time-derivative (recall that the temperature is assumed constant here), we can express the differential equation more simply as a conservation law. 
This is no surprise: the evolution of the chemical potential was originally introduced through the anomaly equation which could be expressed as a conservation of helicity, as discussed in \ref{sec:eqns_of_chiMHD}.
Normally, this equation would still be an integral equation as it involves both $k_\CPI{}(t)$ and its time-average $\overline k_\CPI{}(t)$. 
However, much after the instability had developed, the time-average becomes roughly proportional to the chiral wavenumber, indicating a polynomial behavior (at least on timescales spanning a few orders of magnitude in $t_\CPI{}$).
Analytic arguments (as presented in the main text and section \ref{app:leading_chemical_potential_scaling}) and numerical simulations also confirm that this scaling is close to $k_\CPI{}(t)\propto t^{-1/2}$, thus we make the approximate substitution of $\overline k_\CPI{}\to 2k_\CPI{}$.
This scaling is {\it not} perfect, but the deviation from it is sufficiently weak to allow for this approximation to be taken here.
The resulting relation is simply an algebraic equation for $k_\CPI{}(t)$:
\begin{equation}
    \label{eq:eq_to_solve_for_kA}
    k_\CPI{}(t)-\frac{3\alpha^2S_i}{\pi^2 T^2_i}\sqrt{\frac{\pi\overline\sigma}{2t}} k_\CPI{}(t)\exp\bigg(\frac{2k_\CPI{}^2(t) t}{\overline\sigma}\bigg)\bigg[\frac{\Gamma\big(-\tfrac{1}{2},2k_\CPI{}^2(t)t/\overline\sigma\big)}{\Gamma(-\tfrac{1}{2})}-2\bigg]\simeq k_{\CPI{}}(t_i)\com
\end{equation}
where the integration constant on the right-hand side is approximately the initial chiral wavenumber up to small corrections (formally one finds the corrections at the level of $\sim\alpha^3$).
The incomplete $\Gamma$ function $\Gamma(s,x)$ appears due to the evaluation of the $k$-integral in \eref{eq:kA_conservation_equation_app}.

\begin{figure}
    \centering
    \includegraphics[width=0.7\linewidth]{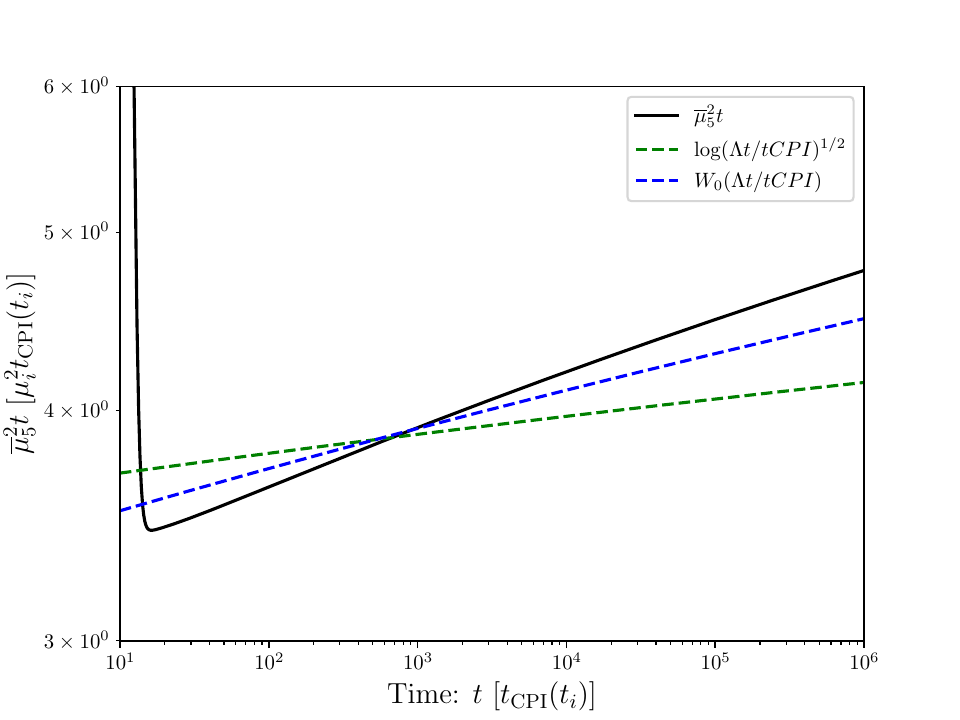}
    \caption{An example time-evolution of the chemical potential compared with analytic scaling formulae. The numerical solution to $\mu_5^2(t)t$, obtained from solving \eref{eq:eqns_of_motion_chromatic}, is shown by the solid black line. For comparison, we have plotted the scaling derived in \eref{eq:W_scaling} and \eref{eq:log_scaling} in green and blue respectively. $\Lambda$ was calculated using \eref{eq:Lambda_def}. This figure shows that these corrections partially capture the increasing $\mu_5^2t$ behavior at late time indicating we have approximated the sub leading behavior of $\mu_5$ at late time.}
    \label{fig:mu_A_polynomial_tilt}
\end{figure}

At this point, solving \eqref{eq:eq_to_solve_for_kA} at late times requires us to make an assumption for how $k_\CPI{}^2(t) t$ behaves.
Clearly, the aforementioned $k_\CPI{}(t)\propto t^{-1/2}$ would lead to a particularly simple relation, however, it does not solve the equation exactly, even in the limit of $t\to\infty$.
Nevertheless, the solution should be close to this function, and thus we are left with two choices for the late-time evolution of $k_\CPI{}(t)$: (i) $k_\CPI{}^2(t)t$ is a {\it growing} function of time or (ii) $k_\CPI{}^2(t)t$ is a {\it decreasing} function of time.
In the latter case, the special function dominates in the square brackets. In the limit of $t\to\infty$, it gives a scaling $k_\CPI{}\sim a-b/t$ which contradicts the assumption so we must constrain ourselves to (i).
In this case the special function is exponentially small at late times and thus we are left with
\begin{equation}
    \label{eq:k5_equation_approx_solvable}
    k_\CPI{}(t)+\frac{6\alpha^2S_i}{\pi^2 T^2_i}\sqrt{\frac{\pi\overline\sigma}{2t}} k_\CPI{}(t)\exp\bigg(\frac{2k_\CPI{}^2(t) t}{\overline\sigma}\bigg)\simeq k_{\CPI{}}(t_i)\per
\end{equation}
This is an implicit equation for $k_\CPI{}(t)$ that is solved by the principal branch of the Lambert $W$ function, as we will show below.
For simplicity, let us look at the following equation:
\begin{equation}   
    \label{eq:lambert_W_algebraic_form}
    y(x) + a\frac{y(x)}{\sqrt{x}}\exp\left(b\,y^2(x)x\right)=c\com
\end{equation}
where $y(x)>0$ and $a,b,c>0$. 
Here, to map from \eref{eq:k5_equation_approx_solvable} to \eref{eq:lambert_W_algebraic_form}, $x\equiv t,~y(x)\equiv k_\CPI{}(t),~a\equiv (6\alpha^2 S_i)/(\pi^2 T_i^2) \sqrt{\pi \overline{\sigma}/2},~b\equiv2/\overline{\sigma}$, and $c\equiv k_\CPI{}(t_i)$.
The solution necessarily satisfies $\lim_{x\to \infty} y(x) = 0^+$, such that at sufficiently large values of $x$, we have $y(x)\ll c$ and we end up with 
\begin{equation}
    \label{eq:Lambert_derivation_1}
    \frac{y(x)}{\sqrt{x}}\exp\left(b\,y^2(x)x\right)\simeq d\com
\end{equation}
where $d=c/a$.
Introducing a new function as $w=2bxy^2$, the square of \eqref{eq:Lambert_derivation_1} leads to
\begin{equation}
    w\exp(w)\simeq 2bd^2x^2\com
\end{equation}
which is the standard form of the implicit defining relation of the Lambert $W$ function.
Since the right-hand side in our case is real and positive, we need only consider the principal branch $W_0(z)$:
\begin{equation} \label{eq:W_scaling}
    y(x) = \sqrt{\frac{W_0(2bd^2x^2)}{2bx}}\per
\end{equation}
We can make the identification among the parameters of \eqref{eq:lambert_W_algebraic_form} and \eqref{eq:k5_equation_approx_solvable} to find the approximate scaling of the chemical potential at late times. 
Using that the Lambert $W_0(z)$ function behaves as $\ln(z)$ for large values of $z$, we find the formal scaling to be:
\begin{equation} \label{eq:log_scaling}
    \mu_5(\tau)\sim \sqrt{\frac{\ln[\Lambda t/t_{\rm CPI}(t_i)]}{t/t_{\rm CPI}(t_i)}}\com
\end{equation}
where  $\Lambda\sim T^2_i/[\alpha^4 S_ik_\CPI{}(t_i)] \gg 1$.

At this point, it is worth asking why the difference from the ``pure'' $\mu_5(t)\propto t^{-1/2}$ scaling is so small.
As we have shown, the smallness of the discrepancy has to do with the vastly different timescales that the problem introduces, \ie{}, the chiral instability timescale (measured in the dimensionless time-like variable $\tau=t/t_\CPI{}(t_i)$)
and the late-time ``drifting'' time-scale (measured in the dimensionless time-like variable $\Lambda\tau$).
The difference between time-scales is characterized by the dimensionless parameter $\Lambda$.
When properly expressed with the physical parameters, it is given as
\begin{equation}
    \label{eq:Lambda_def}
    \Lambda^2\simeq \frac{8 \pi^5 T^4_i}{9\alpha^6 S_i^2\mu_{i}^2}\sim \frac{T^4_i}{\alpha^6 S_i^2\mu_{i}^2}\per
\end{equation}
We can further manipulate this by using the fact that $S_i=\mathcal{S}_{BB}(|\kvec|,t_i)$ is the initial spectrum of the electromagnetic field, and thus it is related to the initial electromagnetic energy density via
\begin{equation}
    \overline\rho_\mathrm{em}(t=t_i)\sim \int _0^{k_\CPI{}(t_i)}{\rm d} k\,k^2\mathcal{S}_{BB}(|\kvec|,t_i)\sim k_\CPI{}^3(t_i) S_i\sim \alpha^3\mu_i^3S_i\com
\end{equation}
where we put a wavenumber cutoff on the initial constant spectrum at the instability scale.
This is a reasonable approximation as most of the energy density is condensed around this wavenumber, and larger modes in the spectrum are exponentially suppressed.
It follows that $S_i\sim \overline\rho_\mathrm{em}(t_i)/[\alpha\mu_{i}]^3$ and thus
\begin{equation}
    \Lambda \sim \frac{\mu^2_iT^2_i}{\overline\rho_\mathrm{em}(t_i)}\sim \frac{\overline\rho_5(t_i)}{\overline\rho_\mathrm{em}(t_i)}\per
\end{equation}
The magnitude of $\Lambda$ is thus a measure of how much initial energy is contained in the chiral asymmetry versus that of the low-$k$ modes of the electromagnetic field.
A similar parameter was found in section~\ref{sec:analytic_sol} when considering the analytic solution of the monochromatic evolution equations, denoted there as $\xi$.
As the monochromatic solution did not have a ``drifting'' region, this parameter only played a small part in the solution (although $\xi\neq0$ was crucial) and the late-time values of the fields were independent of it.

\bibliographystyle{JHEP}
\bibliography{refs}

\end{document}